\documentclass[prd,onecolumn,floatfix,preprintnumbers,letterpaper,amsmath,amssymb,nofootinbib,superscriptaddress]{revtex4}
\usepackage[latin1]{inputenc}
\usepackage{graphicx}
\usepackage{amsmath}
\usepackage{amsfonts}
\usepackage[english]{babel}
\usepackage{amssymb}
\usepackage{multirow}
\usepackage{color}
\usepackage{epsfig}
\usepackage{bm}
\usepackage{subfigure}
\RequirePackage[colorlinks,citecolor=blue,urlcolor=magenta,linkcolor=blue]{hyperref}
\usepackage{url}
 \usepackage[titletoc,toc,title]{appendix}
 
 \def\RN{Reissner-Nordstr\"{o}m }
 \def\KN{Kerr-Newmann }
\begin{document}
	\title{\bf{Astrophysical Signatures of Black holes in Generalized Proca Theories}}
	\author{Mostafizur Rahman}
\email{mostafizur@ctp-jamia.res.in}
\affiliation{Centre for Theoretical Physics, Jamia Millia Islamia, New Delhi-110025, India.}
\author{Anjan A. Sen }
\email{aasen@jmi.ac.in}
\affiliation{Centre for Theoretical Physics, Jamia Millia Islamia, New Delhi-110025, India.}
	\begin{abstract}
	\noindent
Explaining the late time acceleration is one of the most challenging tasks for theoretical physicists today. Infra-red modification of Einstein's general theory of relativity (GR) is a possible route to model late time acceleration. In this regard, vector-tensor theory as a part of gravitational interactions on large cosmological scales, has been proposed recently. This involves generalization of massive Proca lagrangian in curved space time. Black hole solutions in such theories have also been constructed. In this paper, we study different astrophysical signatures of such black holes. We first study the strong lensing and time delay effect of such static spherically symmetric black hole solutions, in particular for the  case of gravitational lensing of the star S2 by Sagittarius A* at the centre of Milky Way. We also construct the rotating black hole solution from this static spherically symmetric solution in Proca theories using the Newman-Janis algorithm and subsequently study lensing, time delay and black hole shadow effect in this rotating black hole space time. We discuss the possibility of detecting Proca hair in future observations.
	\end{abstract}
	\maketitle

	\section{Introduction}\label{intro}

Einstein's General Theory of Relativity (GR) is an extremely successful theory to describe gravity from Solar System scales involving planetary motions upto Cosmological scales describing the expansion of the Universe, formation of light elements, existence of cosmic microwave background radiation, formation of large scale structures. But the late time acceleration of the Universe, first confirmed by SNIa observations two decades ago \cite{Riess1998, Perlmutter1999, Tonry2003, Knop2003, Riess2004}, is the first observed astrophysical phenomena, that attractive gravity fails to explain. Accelerated expansion in the Universe demands the existence of repulsive gravity at large cosmological scales. This can be done if one modifies either the matter part with exotic components having negative pressure, or modifies the gravity at large cosmological scales (see \cite{copeland, Padmanabhan, Peebles, Sahni}, for review on this topic). Although the cosmological constant ($\Lambda$), as introduced by Einstein to model a static Universe, is the simplest solution to the late time acceleration of the Universe, the large discrepancy between the observed value of $\Lambda$ and what we expect its value from field theory point of view, is the greatest obstacle for it to be a successful explanation for the late time acceleration of the Universe ( also recent observations suggest tensions in $\Lambda$CDM with the data \cite{R16, R18}). A consistent theory of quantum gravity is needed to solve this cosmological constant problem.
	
Going beyond $\Lambda$, whether to modify the matter sector or the gravity sector, scalar fields play the most important role in late time acceleration of Universe \cite{copeland}. Scalar fields do exist in nature; Higgs field, which is the fundamental ingredient of standard model of particle physics \cite{higgs}, is the best example of a scalar field that exists in nature. Moreover, being a scalar, it can be naturally incorporated in a isotropic and homogeneous Universe. It also can give rise to repulsive gravity with its slow-roll property and hence can explain late time acceleration. But these scalar fields have to be very light in order to slow-roll at large cosmological scales and without any mechanism to avoid their possible interactions with baryons, they give rise to long-range fifth force in baryons that is absent in solar system scales. To avoid such tensions, we need to have some screening mechanism that prevents the scalar field to interact with baryons on small scales, but allows the scalar field to give desired late time accelerated expansion at large cosmological scales. Chameleon mechanism \cite{chameleon}, Vainshtein mechanism \cite{vainstein} are examples of such screening processes.

Amongst the scalar field models for infra-red modification of gravity, Galileon model is one of the most studied models \cite{galileon, covariant, lategalileon}. It was first introduced as a natural extension of DGP brane-world model \cite{dgp} in decoupling limit \cite{decoup}. The lagrangian for the Galileon field respects the shift symmetry and contains higher derivative terms. Despite this, the equation of motion for Galileon field is second order and hence the theory is free from Ostragradsky ghosts \cite{ostro}. One can also implement the Vainshtein mechanism in this model to preserve the local physics and to satisfy the solar system constraints. The general Galileon action with second order equations of motion, contains non-minimal derivative coupling with Ricci and Einstein tensor. This is a subclass of more general Horndeski theories \cite{horndeski} which contain scalar-tensor interactions with second order equation of motion on curved background. Massive gravity theories \cite{massive, galileon} are other examples of general scalar-tensor theories giving second order of equations of motion.

Similar to scalar-tensor theories, one can also have consistent models of vector-tensor theory as a part of the gravitational interactions on large scales resulting the late time acceleration in the Universe \cite{Tasinato:2014eka}. In Minkowski space, allowing the mass of the vector fields, leads to Proca lagrangian. One can then generalize this Massive Proca lagrangian to curved space time. This has been done in a recent paper by Heisenberg \cite{Heisenberg:2014rta}, where a generalized massive Proca lagrangian in curved background with second order equations of motion has been proposed. This constitutes a Galileon type self interaction for the vector field including the non-minimal derivative coupling to gravity. Different cosmological aspects of such models as well as constraints from cosmological observations have been studied in several recent works \cite{cosmoproca}. In a recent paper, Heisenberg has studied, in systematic way, different generalisations of Einstein gravity and their cosmological implications \cite{heisen_recent} 

The recent results from Advanced Ligo experiment for measuring gravitational waves \cite{ligo}, have opened up the opportunity to probe astrophysical black holes. The latest gravitational wave measurements from two colliding neutron stars and its electromagnetic counterparts \cite{ligoneutron}, have confirmed the validity of GR for these astrophysical processes. This put extremely tight constraints on different modified gravity theories based on scalar fields \cite{ligode}, explaining the late time acceleration in the Universe. In a recent work, Jimenez and Heisenberg \cite{Jimenez:2016isa} have put forward vector models for dark energy based on Proca lagrangian with $c_{gw} =1$ making it consistent with latest Ligo observations for neutron star merger. But the model can still give non-trivial predictions for gravitational waves.

To probe any gravitational theory at astrophysical scales, black hole are the best candidates. Recently, Heisenberg et al. have constructed hairy black hole solutions in generalized Proca theories \cite{Heisenberg:2017xda}. For power-law coupling, they found a class of asymptotically flat hairy black hole solutions. These are not exact solutions but are iterative series solutions upto $\mathcal{O}(1/r^3)$ which matches excellently with the numerical solutions. These are hairy black hole solution in a modified gravity scenario and it is extremely interesting to study their astrophysical signatures to probe the underlying modified gravity theory.

Gravitational lensing is one of most interesting astrophysical phenomena due to gravitational effects of massive bodies. It is broadly the bending of light due to the curvature of the space time and as the curvature of the space time depends on the gravitational properties of massive bodies, one can directly constrain different properties of a massive body like its mass or angular momentum, by observing its gravitational lensing effect. In solar system, through lensing effects, observers first confirmed the validity of Einstein GR. But in solar system, the effect is pretty weak with deflection angle much small compared to $2\pi$ \cite{lense1,lense2,lense3}. But it can be large in the vicinity of strong gravitating objects like black holes, where the photon can circle in closed loops around the black hole many times due to the strong gravitational effect, before escaping. There exists a sphere around the black hole called "{\it photon sphere}", where the deflection angle for the photon can even diverge. Gravitational lensing in the space time of Schwarzschild black holes was first studied by Virbhadra and Ellis \cite{virbhadra} and later it was extended to Reissner-Nordstorm \cite{lensrn} and Kerr black holes \cite{lenskerr}, black holes in  brane-world models \cite{lensbrane} and Galileon models \cite{lensgal}, in extra dimension with Kalb-Ramond field \cite{lenskalb} and so on. As strong gravitational lensing in the vicinity of black holes probes different properties of the black holes, it is also useful to probe different modified gravity theories as  standard Schwarzschild or Kerr black solutions get modified in different versions of modified gravity theories. Moreover through gravitational lensing, one can probe the region around black holes, known  as ``{\it black hole shadow}" \cite{shadow}. The shape and size of the black hole shadow is a direct probe for the black hole space time and hence the underlining gravity theory. With the prospects of Even Horizon telescope \cite{eht} as well as telescopes like SKA \cite{ska}, one can resolve the black hole shadow with great accuracy and hence probing modified gravity through such observations is possible in near future.

In this paper, we study the strong lensing phenomena for the black hole spacetimes in generalized Proca theories. Throughout the paper, we have used the geometrical unit $G=c=1$.

\section{Hairy black hole solution in generalized Proca theories}\label{sec2}

The action for general Proca theory is given by  \cite{Heisenberg:2017xda, Heisenberg:2014rta, Jimenez:2016isa, Heisenberg:2017hwb}:
\begin{equation}{\label{intro1}}
S=\int d^4x \sqrt{-g} \biggl( F+\sum_{i=2}^{6}{\cal L}_i 
\biggr)\,,
\end{equation}
with
\begin{eqnarray}{\label{intro2}}
& &{\mathcal L}_2=G_2(X)\,,\qquad 
{\mathcal L}_3=G_3(X) {\mathcal A^{\mu}}_{;\mu}\,, \nonumber \\
& &{\mathcal L}_4=G_4(X)R+G_{4,X} \left[ 
({\mathcal A^{\mu}}_{;\mu})^2 -{\mathcal A_{\nu}}_{;\mu} {\mathcal A^{\mu}}^{;\nu} 
\right]-2g_4(X)F\,, \nonumber \\
& &{\mathcal L}_5=G_5(X)G_{\mu \nu} {\mathcal A^{\nu}}^{;\mu}
-\frac{G_{5,X}}{6} [ ({\mathcal A^{\mu}}_{;\mu})^3
-3{\mathcal A^{\mu}}_{;\mu} {\mathcal A_{\sigma}}_{;\rho}
{\mathcal A^{\rho}}^{;\sigma} \nonumber \\
& &~~~~~+2{\mathcal A_{\sigma}}_{;\rho}{\mathcal A^{\rho}}^{;\nu}
{\mathcal A_{\nu}}^{;\sigma}]
-g_5(X)\tilde{F}^{\alpha \mu} {\tilde{F}^{\beta}}_{\mu} 
\mathcal A_{\beta;\alpha}\,,\nonumber \\
& &{\mathcal L}_6=G_6(X) L^{\mu \nu \alpha \beta}
\mathcal A_{\nu;\mu}\mathcal A_{\beta;\alpha}
+\frac{G_{6,X}}{2}\tilde{F}^{\alpha \beta} {\tilde{F}^{\mu \nu}} 
\mathcal A_{\mu;\alpha}\mathcal A_{\nu;\beta}.
\end{eqnarray}
Here $F=-F_{\mu \nu}F^{\mu \nu}/4$. The functions $G_{2} - G_{6}$ as well as $g_{4}$ and $g_{5}$ depend on 
$X=-\mathcal{A}_{\mu}\mathcal{A}^{\mu}/2 $. We denote 
$G_{i,X}=\partial G_i/\partial X$. The vector field $\cal A^{\mu}$ has 
non-minimal couplings with space time curvature through
$L^{\mu \nu \alpha \beta}={\cal E}^{\mu \nu \rho \sigma}
{\cal E}^{\alpha \beta \gamma \delta}R_{\rho \sigma \gamma \delta}/4$, 
where ${\cal E}^{\mu \nu \rho \sigma}$ is the 
Levi-Civita tensor 
and $R_{\rho \sigma \gamma \delta}$ is the Riemann tensor. 
The dual strength tensor  $\tilde{F}^{\mu \nu}={\cal E}^{\mu \nu \alpha \beta}F_{\alpha \beta}/2$.
The Einstein-Hilbert term $M_{\rm pl}^2/2$ is contained in $G_4(X)$.

To describe the black-holes in this general Proca theory, one assumes a static spherically symmetric space time:

\begin{equation}{\label{intro3}}
ds^{2} =-A(r) dt^{2} +B(r)dr^{2} + 
C(r) \left( d\vartheta^{2}+\sin^{2}\vartheta\,d\varphi^{2} 
\right)\,,
\end{equation}

\noindent
together with the vector field $\mathcal{A}_{\mu}=(\mathcal{A}_{0}(r),\mathcal{A}_{1}(r),0,0)$. Here $A(r)$, $B(r)$, $\mathcal A_0(r)$, and $\mathcal A_1(r)$ are arbitrary functions of $r$. In  \cite{Heisenberg:2017xda, Heisenberg:2017hwb},  the following action has been considered for general Proca theory:

\begin{equation}{\label{intro4}}
S=\int d^4x \sqrt{-g} \biggl( \frac{M_{Pl}^2}{2}R +\beta_{3} {\cal A}^{\mu}_{;\mu} X + F\biggr).
\end{equation}

%For power-law coupling, one assumes \cite{Heisenberg:2017xda} :

%\begin{equation}{\label{intro4}}
%G_i(X)=\tilde{\beta}_i X^n\,,\qquad 
%g_j(X)=\tilde{\gamma}_{j}X^n\,,
%\end{equation}
%
%where $n$ is a positive integer, and $\tilde{\beta_i}$ and 
%$\tilde{\gamma}_j$ are coupling constants. Here $i=3,4,5,6$ and $j=4,5$. In this case, an asymptotic black hole solution %was found with iterative method which agrees pretty well with the exact numerical solution. 

Upto $\mathcal{O}(1/r^3)$, the Black hole solution for such theory is given by \cite{Heisenberg:2017xda, Heisenberg:2017hwb} :

\begin{equation}{\label{intro5}}
	\begin{aligned}
	A(r)&=1-\frac{2}{r}-\frac{P^{2}}{6r^{3}} + \mathcal{O}(1/r^4)\\
	B(r)^{-1}&=1-\frac{2}{r}-\frac{P^{2}}{2r^{2}}-\frac{P^{2}}{2r^{3}} + \mathcal{O}(1/r^4)\\C(r)&=r^{2}
	\end{aligned}
	\end{equation}
	where, we have set $r=r/M$, where $M$ is the mass of the black hole. Throughout the paper, all the distances are measured in the unit of the mass of the black hole ($M=1$) unless otherwise specified. Here $P$ is Proca hair, related to the time component of the vector field as $\mathcal{A}_0=(P-P/r-P/(2r^2))M_{Pl}+{\cal O}(1/r^3)$. We set $P=P/M_{Pl}$ where $M_{Pl}$ is the Planck mass. Clearly, the metric satisfies asymptotically flat condition, $\lim\limits_{r \to \infty}A(r)=\lim\limits_{r \to \infty}B(r)=1 $. Note that, in the limit $P\to 0$ i.e. when the Proca hair $P$ vanishes, the above metric elements reduce to that of Schwarzschild metric.

	\section{Lensing Effect in Strong Field Limit in a Static, Spherically symmetric metric}
	Before considering the spacetime of our interest, we review the gravitational lensing effect in Strong Field Limit (SFL) in a general asymptotically flat, static and spherically symmetric space-time. In this section we discuss about the main concepts and different observables related to gravitational lensing in the strong field limit following Ref. \cite{Bozza:2002zj}. 
	
\subsection{Observables in Strong Field Limit}\label{bb}
\begin{figure}
					\includegraphics[width=\linewidth]{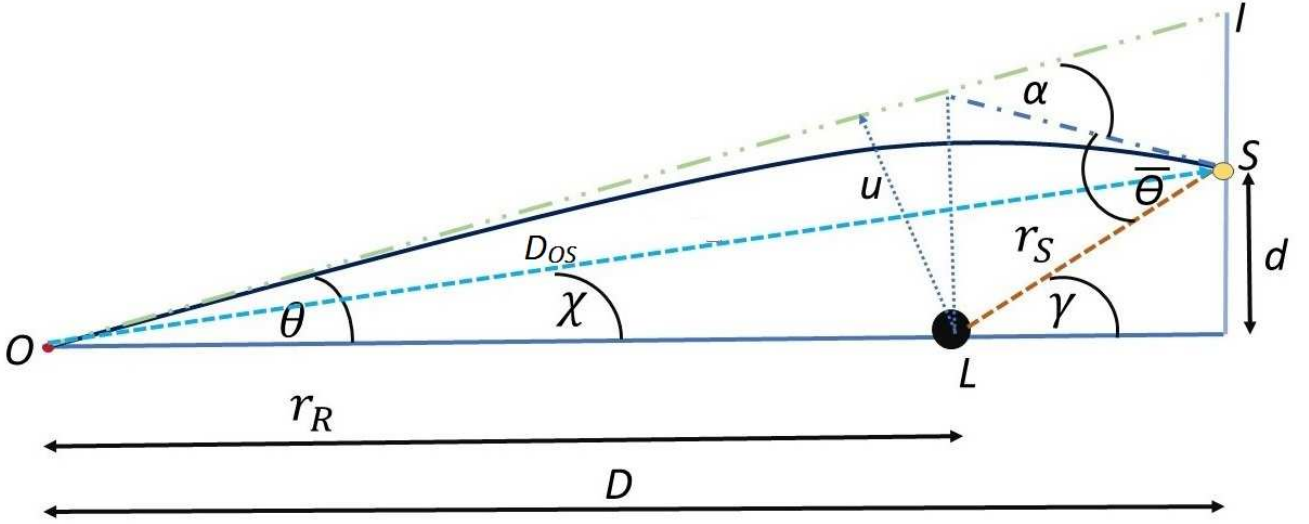}
					\caption{A schematic diagram of the lensing system has been presented. Light from source S get lensed by the black hole L and incident on the observer O with an angle $\theta$. Image is formed at I. The line joining O and L is called the Optic axis \cite{lenskalb}. }
						\label{fig:Lensing system}
			\end{figure}

	 Any generic static, spherically symmetric space-time can be described by the line element (\ref{intro3}).
 In order to study the photon trajectory, we will assume that the equation \cite{Claudel:2000yi, virbhadra, Bozza:2002zj}
 	\begin{equation}{\label{2}}
        C'(r)A(r)-A'(r)C(r)=0
		\end{equation}
		admits at least one positive solution and the largest positive solution of this equation is defined as the radius of photon sphere, $r_{m}$. We further assume that $A(r)$, $B(r)$ and $C(r)$ are finite and positive for  $r\geq r_{m}$ \cite{Bozza:2002zj}. Since the spacetime admits spherical symmetry, we can restrict our attention to equatorial plane ($\vartheta=\pi/2$) without losing any generality. Now we can formulate the lensing problem. Consider a black hole situated at the origin. A photon with impact parameter $u$ incoming from a source situated at  $r_{S}$, deviates while approaching it. Let the photon approaches the black hole at a minimum distance $r_{0}$ and then deviated away from it. An observer situated at $r_{R}$ detects the photon (see Fig.-(\ref{fig:Lensing system})). In the strong field limit, we consider only those photons whose closest approach distance $r_{0}$ is very near to $r_{m}$ and hence the deflection angle $\alpha$ can be expanded around the photon sphere, $r_{m}$ or equivalently minimum impact parameter $u_{m}$. When the closest approach distance $r$ is greater than $r_{m}$, it just simply gets deflected (it may complete several loops around the black hole before reaching the observer). When it reaches a critical value $r_{0}=r_{m}$ (or $u=u_{m}$), $\alpha$ diverges and the photon gets captured. Following the method developed by Bozza \cite{Bozza:2002zj}, one can show that this divergence is logarithmic in nature and the deflection angle can be written as 
		\begin{equation}{\label{6}}
		\alpha(\theta)=-\bar{a} \ln\left(\frac{\theta}{ \theta_{m}}-1\right)+\bar{b}
		\end{equation}
		where subscript $`m$' denotes function evaluated at $r=r_{m}$. $\theta$ is the incident angle to the observer whereas $\theta_{m}=u_{m}\sqrt{A(r_{R})/C(r_{R})}$ corresponds to the incoming
		photon with minimum impact factor, $u_{m}=\sqrt{C_{m}/A_{m}}$. When $\theta\leq\theta_{m}$, the photon gets captured. The parameters $\bar{a}$ and $\bar{b}$ are called the Strong Lensing coefficients whose functional forms are given in Eq. (35-36) of Ref.\cite{Bozza:2002zj}. 
			
			With the help of Eq. (\ref{6}), we can calculate the observables for strong lensing corresponding to any given static and spherically symmetric metric using the lens equation. The corresponding observables are - (i) position of the innermost image, $\theta_{m}$ ,(ii) the angular separation between the first relativistic image (outermost image) with the innermost image, $s$, and (iii) relative flux between different images,  $\mathcal{R}$. In order to do so, we introduce co-ordinate independent lens equation: $\alpha=\theta-\theta_{S}+\phi_{RS}$, where $\theta$ and $\theta_{S}$ denote the angles that are measured at the receiver position and the source position, respectively, while $\phi_{RS}$ is the angle between the azimuthal coordinate of source and observer \cite{Ishihara:2016vdc}. The quantity $\alpha$ is geometrically invariant which in asymptotically flat limit, coincides with the deflection angle.	If $\bar{\theta}$ denotes the impact angle as seen from the  source, then angle measured from source position becomes  $\theta_{S}=\pi-\bar{\theta}$. Let $\gamma$ be the angle between the optic axis (the line joining the observer and the lens) and the line joining the lens and the source (see Fig-\ref{fig:Lensing system}). Note that $	\phi_{RS}=\pi -\gamma$. Then the lens equation connecting the observer and source position takes the form
				\begin{equation}{\label{8}}
\gamma=\theta+\bar{\theta}-\alpha(\theta)
				\end{equation}
	This equation is known as the Ohanian lens equation \cite{1987AmJPh..55..428O} and as discussed in Ref.\cite{Bozza:2008ev}, is the best approximate lens equation in asymptotically flat space-time. Since both the source and observaer is situated far away from the black hole, $\bar{\theta}$ can be approximated as $\bar{\theta}=\theta r_{R}/r_{S}$. With this condition and using Eq. (\ref{6}) and (\ref{8}), one can obtain the position of the $n$-th order image \cite{Bozza:2002zj}
			\begin{equation}{\label{9}}
			\theta_{n}= \theta_{m}\left(1+\exp\left(\dfrac{\bar{b}+\gamma-2n\pi}{\bar{a}}\right)\right) 
			\end{equation} Where $n$ corresponds to the number of winding around the black hole.  When $n\to\infty$, $\theta_{n}$ becomes $\theta_{m}$. So $\theta_{m}$ represents the position of inner most relativistic image. In simplest of situation, we consider that the outermost image (first relativistic image) $\theta_{1}$ is resolved as a single image and all the other images packed together at $\theta_{m}$ \cite{Bozza:2002zj, lenskalb}. Then the angular separation between these two images is defined as \cite{Bozza:2002zj}
			\begin{equation}{\label{10}}
			s= \theta_{1}-\theta_{m}=\theta_{m} \exp\left(\dfrac{\bar{b}+\gamma-2\pi}{\bar{a}}\right)
			\end{equation}

Magnification of the image is defined as the ratio of solid angle to the observer with a lens to the solid angle without lens i.e. $	\mu=\sin\theta d\theta/\sin\chi d\chi$, where $\chi$ is the angle between source to observer w.r.t. the optic axes (see Fig-\ref{fig:Lensing system}). Note that lens equation Eq. (\ref{8}) does not have any term that contains $\chi$. So using the relation, $	r_{S}\sin\gamma=D_{OS}\sin\chi$ and considering $D_{OS}\gg r_{S}$, one can easily show that the magnification of $n$th relativistic image can be written as \cite{Bozza:2002zj}
\begin{eqnarray}{\label{11}}
& &{\mu}_{n}=\left(\frac{D_{OS}}{r_{S}}\right)^{2}\dfrac{\theta_{m}^{2} e_{n}(1+e_{n})}{\bar{a}\sin\gamma}\,,\qquad 
{e}_{n}=\exp \left(\dfrac{\bar{b}+\gamma-2n\pi}{\bar{a}}\right)\,. \nonumber 
\end{eqnarray}

where $D_{OS}$ is the distance between the source and observer. The ratio of magnification hence the flux from the first relativistic image to all the other images is given by \cite{Bozza:2002zj}
			\begin{equation}{\label{13}}
			\mathcal{R}=2.5\log_{10}\left(\dfrac{\mu_{1}}{\sum\limits_{n=2}^{\infty}\mu_{n}}\right)=\frac{5\pi}{\bar{a}\ln 10}
			\end{equation}
			If we have a precise knowledge of $\gamma$ and observer to lens distance $r_{R}$, then we can predict strong lensing co-efficient $\bar{a}$, $\bar{b}$ and minimum impact parameter $u_{m}$ by measuring $\mathcal{R}$, $s$, $\theta_{m}$. Then by comparing them with the values predicted by given theoretical models, we can identify the nature of the black hole.	
\subsection{Time delay in strong field gravitational lensing}\label{Td} 
	In this section we briefly review the Time Delay effect in a static, spherically symmetric spacetime following the method developed by Bozza and Manchini \cite{Bozza:2003cp}. From the discussion in the previous section, it is clear that formation of multiple images is a key feature of strong lensing and generally the time taken by different photons following different  paths (which correspond to different images) are not the same. So there are some time delay between different images. Moreover time delay between the images will depend on which side of the lens the images are formed. When both the images are on the same side of the lens, time delay between $m$ and $n$ th relativistic image can be expressed as \cite{Bozza:2003cp} 
			 \begin{equation}{\label{Td1}}
			 \Delta T_{mn}^{s}=-u_{m} 2\pi(m-n)+2\sqrt{u_{m}}\sqrt{\frac{B_{m}}{A_{m}}}\left(\exp(\dfrac{\bar{b}+\gamma-2m\pi}{2\bar{a}})-\exp(\dfrac{\bar{b}+\gamma-2n\pi}{2\bar{a}})\right)
			 \end{equation} The sign of $\gamma$ depends on which side of the source images are formed. When the images are formed on the opposite side of the lens time delay between $m$ and $n$ th relativistic image can be expressed as \cite{Bozza:2003cp}
			  \begin{equation}{\label{Td2}}
			  \Delta T_{mn}^{o}=-u_{m} (2\pi(m-n)-2\gamma)+2\sqrt{u_{m}}\sqrt{\frac{B_{m}}{A_{m}}}\left(\exp(\dfrac{\bar{b}+\gamma-2m\pi}{2\bar{a}})-\exp(\dfrac{\bar{b}-\gamma-2n\pi}{2\bar{a}})\right)
			  \end{equation}
			We need instruments with high observational precision in order to find the contribution from the 
			 second term.
			 Thus for practical purposes, we can  approximate the time delay by its first term's contribution. In terms of $\theta_{m}$ one can get an interesting result when both the images are formed in the same side of the lens. Then time delay between first and second relativistic image can be expressed as \cite{Bozza:2004kq}
			 \begin{equation}{\label{Td3}}
			 \Delta T_{12}^{s}=\theta_{m}2\pi r_{R}
			 \end{equation}
			 In principle, using this formula, we can get a very accurate estimate for the distance of the black hole. Note that for a distant observer, $A(r_{R})$ practically becomes 1 and $\theta_{m}$ can be written as $\theta_{m}=r_{m}/\sqrt{A_{m}r_{R}^{2}}$. Using Eq. (\ref{Td3}), we can found a interesting result given by \begin{equation}{\label{Td4}}
	r_{m}^{2}A(r_{m})=\left(\dfrac{\Delta T_{12}^{s}}{2\pi}\right)^{2}
	\end{equation}
	 This equation beautifully relates an observational parameter, the time delay $\Delta T_{12}^{s}$ between first and second relativistic image with a theoretical parameter, the metric function $A(r)$ evaluated at $r=r_{m}$. Thus one can verify a given theoretical model by solving this equation using the observational data of $\Delta T_{12}^{s}$.
	 \begin{table}
			 	 	\begin{center}
			 	 		\def\arraystretch{1.5}
			 	 		\setlength{\tabcolsep}{1.5em}
			 	 		\begin{tabular}{|c|c|c|c|c|c|c|c|}
			 	 			\hline  
			 	 			Orbital Parameter&$\varrho$ (pc)& $T$ (yr)&$e$&$T_{o}$ (yr)&$i$ (deg)&$\Omega$ (deg)&$\omega$ (deg) \\ \hline
			 	 			Value&$4.54\times 10^{-3}$&$15.92$&$ 0.89$&$ 2018.37$&$ 45.7$&$ 45.9$&$244.7$ \\
			 	 			\hline
			 	 		\end{tabular}
			 	 	\end{center}
			 	 		\caption{Orbital parameters for S2. Its orbit can be described by an ellipse with  $\varrho$ is the semi major axis and $e$ being the eccentricity of the orbit. The inclination angle $i$ denotes the angle between the ellipse and a reference plane in the line of sight. $\Omega$ and $\omega$ describes the position angle of the ascending node and the periapse anomaly with respect to the ascending node respectively. The orbital time period is described by $T$. $T_{0}$ describes the epoch when it reaches the periapse position \cite{Hees:2017aal, Bozza:2004kq}.}\label{tab1}
			 	 \end{table}
\section{Numerical estimation of different observables for Gravitational lensing of the star S2 by Sgr A*}
In this section we will numerically estimate the values of different observable parameters related to strong lensing for a spacetime described by Eq. (\ref{intro3}-\ref{intro5}). For this purpose, we take the nature of the super massive black hole (SMBH) at the center of our galaxy (Sgr A*) is given by  solutions of second order generalized Proca theories. Here we take the star S2 as the source. This star revolves around the SMBH in a highly elliptic orbit with orbital time period around 15.92 years and has the minimum average distance from it. In the early 2018, it had been at its periapse position. Previous studies have shown that the magnification of images is maximum when the star reaches its pariapse position \cite{Bozza:2004kq}. This gives us an unique opportunity to observe different lensing parameters in this time and thus make it possible to verify different theories of gravity. In this section, we first reconstruct the lensing system using the data given in Ref. \cite{Hees:2017aal, Bozza:2004kq} and then numerically calculate different observables related to strong lensing in this scenario.	
\subsection{The Lensing system}
			 
\begin{figure}
			 	 	\centering 
			 	 	\begin{minipage}[b]{0.475\textwidth}
			 	 		\includegraphics[width=\textwidth]{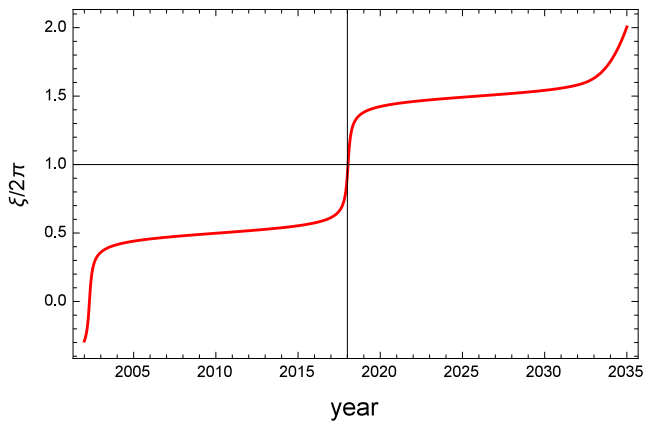}
			 	 	\end{minipage}
			 	 	\hfill
			 	 		\caption{Orbital position of S2 as a function of time have been presented. Here $\xi$ represents the  anamoly angle from the periapse position. Previous studies have shown that the maximum magnification of the images will be obtained when S2 is in its periapse position i.e. when $\xi=2 \kappa \pi$, where $\kappa$ is an integer \cite{Bozza:2004kq}. The plot indicates that this had been achived in early 2018.  }\label{fig:positio}
			 	 \end{figure}

The mass of black hole at the center of our galaxy is estimated to be $4.01*10^{6}M_{\bigodot}$ which is located at a distance $7.8$ kpc away from us \cite{2016ApJ...830...17B}. S2 is one of the star with the minimum average distance from it (S-102 has even smaller minimum average distance but it is 16 times fainter than S2 \cite{S2}). As stated earlier, it was at its periapse position in early 2018 where one expect to have maximum magnification \cite{Bozza:2004kq}. So we have used it as a source for gravitational lensing. Moreover, S2 has radius of few solar radii, so one can treat it as point source. It's orbital motion (along with other short-period stars around SMBH) has been studied over 20 years mainly by two groups, one at Keck Observatory while the other with New Technology Telescope (NTT) and with Very Large Telescope (VTT) \cite{Hees:2017aal}. From those studies, we now have precise understanding about its orbital motion. Its orbital parameters  are reported in Table-\ref{tab1} \cite{Hees:2017aal, Bozza:2004kq}.   It's position ($r_{S},\gamma$), can be expressed in terms of the orbital parameters of the system\cite{Bozza:2004kq, BinNun:2010se}
 \begin{align}{\label{17}}
 r_{S}&=\dfrac{\varrho(1-e^{2})}{1+e\cos\xi}\\r_{R}&\simeq D_{os}=7.8 kpc\\\gamma&=\arccos[\sin(\xi+\omega)\sin i]
\end{align}
where $ D_{os}$ is the distance between observer and the source,  $\varrho$ is the major semi axis, $e$ is the
			 	 eccentricity, $i$ is the inclination of the normal of the orbit with
			 	 respect to the line of sight, $\omega$  is the periapse anomaly with respect to the ascending node. $\xi$ is the anomaly
			 	 angle from the periapse, determined by the differential equation and initial condition 
	 		\begin{equation}\label{20}
	 		\begin{aligned}
	 	\frac{T}{2\pi}\dfrac{(1-e^{2})^{3/2}}{(1+e\cos\xi)^{2}}\dot{\xi}=1\\
\xi(T_{0})=2 \kappa \pi
	 		\end{aligned}
	 		\end{equation}
			 	 where $T$ is the orbital time period of S2 and $T_{0}$ is the epoch of periapse and $\kappa$ be any integer. We have plotted anomaly angle as a function of time in Fig--\ref{fig:positio}. From the plot we can see that the star reached its periapse position in early 2018. 
	\subsection{Numerical Estimation of Different Lensing Parameters}	
	In this section, we present the numerical estimation of different observational parameters considering the SMBH at the center of our galaxy as a lens and the star S2 as a source. Here we  have considered that nature of black hole space-time is given by solutions of second order generalized Proca theories presented in Eq. (\ref{intro5}) and the S2 star is at its pariapse position. 
 The radius of the photon sphere is given by the largest positive solution of the equation (see Eq. (\ref{2}))
	\begin{equation}{\label{23}}
		12r^{3}-36r^{2}-5P^{2}=0
	\end{equation}
	Clearly, one can see that in the limit $P=0$, the radius of the photon sphere is reduces to $r_{m}=3$, representing photon circular orbit in Schwarzschild space-time. By solving the above equation, one can express the radius of the photon sphere as 
	\begin{equation}
	r_{m}=1+\frac{2}{\sqrt[3]{K1}}+\frac{\sqrt[3]{K1}}{2}
	\end{equation} where,
	\begin{equation}
	K1\nonumber=\left[{\left(\frac{5 P^2}{3}+8\right)+\frac{\sqrt{5}}{3} \sqrt{P^2 \left(5 P^2+48\right)}}\right]
\end{equation}	 As stated earlier, we assumed that $A(r)$, $B(r)$,
	and $C(r)$ are finite and positive for  $r\geq r_{m}$. But here $B(r)$ fails to remain positive for $P\geq2.48$ at $r=r_{m}$ and hence in our analysis we will concentrate in the range $P<2.48$. In Table-\ref{tab:tab2}, we have presented the numerical estimation of different observational parameters namely the angular position of the inner most image $\theta_{m}$, the angular separation  between inner and outermost image $s$, the relative magnification of the outermost relativistic image with the other images $\mathcal{R}$ and the time delay between first and second relativistic image $\Delta T^{s}_{12}$ (formed on the same side of the lens). We also compare the results with those obtained from Reissner-Nordstr\"{o}m (RN) black hole solution with charge $q$  whose line element can be expressed as \cite{1916AnP...355..106R, chandra}
%%%%%%%%%%%%%%%%%%%%%%%%%%%%%%%%%%%%%%%%%%%%%%	
	\begin{equation}\label{RN}
	ds^{2}=-\left(1-\frac{2M}{r}+\frac{q^{2}}{r^{2}}\right)dt^{2}+\left(1-\frac{2M}{r}+\frac{q^{2}}{r^{2}}\right)^{-1}dr^{2}+r^{2}\left(d\theta^{2}+\sin^{2}\theta d\phi^{2}\right)
	\end{equation}
%%%%%%%%%%%%%%%%%%%%%%%%%%%%%%%%%%%%%%%%%%%%%%%%%%%%%%%%%%%%%%%%%%%%%%%%%%%%%%%%%%%%%%%%%%%%%%%	
		\begin{table}[]
			\centering
			      	\def\arraystretch{.7}
			      	\setlength{\tabcolsep}{1.3em}
			\begin{tabular}{|l|l|l|l|l|l|l|l|l|}\hline
				\multirow{2}{*}{$hair$} & \multicolumn{2}{l|}{$\theta_{m}$ in $\mu$as} & \multicolumn{2}{l|}{$s$ in $\mu$as} & \multicolumn{2}{l|}{$\mathcal{R}$} & \multicolumn{2}{l|}{$\Delta T_{12}^{s}$ in sec.} \\\cline{2-9} 
				& Proca BH       & RN BH          & Proca BH     & RN BH       & Proca BH     & RN BH        & Proca BH         & RN BH           \\\hline
				0                  & 19.0033     & 19.0033     & 0.182214  & 0.182214  & 15.708     & 15.708     & 32.6484        & 32.6484       \\
				0.3                & 19.0191     & 18.713      & 0.185641  & 0.188735  & 15.5854    & 15.5434    & 32.6755        & 32.1498       \\
				0.6                & 19.066      & 17.7691     & 0.196129  & 0.214982  & 15.2132    & 14.934     & 32.7563        & 30.5281       \\
				0.9                & 19.143      & 15.7962     & 0.214176  & 0.314308  & 14.5764    & 13.0677    & 32.8885        & 27.1385   \\ \hline  
			\end{tabular}
				\caption{Numerical estimations of the observables related to strong lensing ($\theta_{m}$, $s$, $\mathcal{R}$, $\Delta T_{12}^{s}$) have been presented. A comparison between the values of the observables obtained from generalized Proca theories ( Proca BH) to those obtained from \RN black hole ( RN BH) have also been presented. Here the parameter `$hair$' corresponds to Proca hair $P$ in the case of Proca BH and charged hair $q$ in the case of RN black holes. Note that, $hair=0$ case corresponds to Schwarzschild black hole. Here the SMBH at the center of our galaxy is taken as the lens whereas the star S2 is taken as the source. The observables have been calculated at the epoch of periapse of the star S2 (early 2018).}
				\label{tab:tab2}
		\end{table}
%%%%%%%%%%%%%%%%%%%%%%%%%%%%%%%%%%%%%%%%%%%%%%%%%%%%%%%%%%%%%%%%%%%%%%%%%%%%%%%%%%%%%%%%%%%%%%%%		
			\begin{figure}
		\centering 
		\begin{minipage}[b]{0.45\textwidth}
			\includegraphics[width=\textwidth]{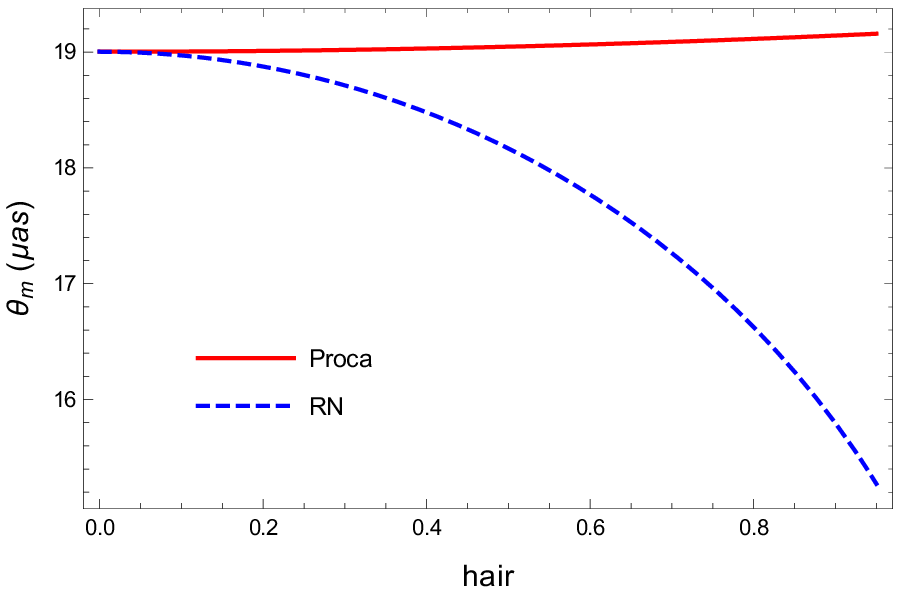}
		\end{minipage}
		\hfill
		\begin{minipage}[b]{0.45\textwidth}
			\includegraphics[width=\textwidth]{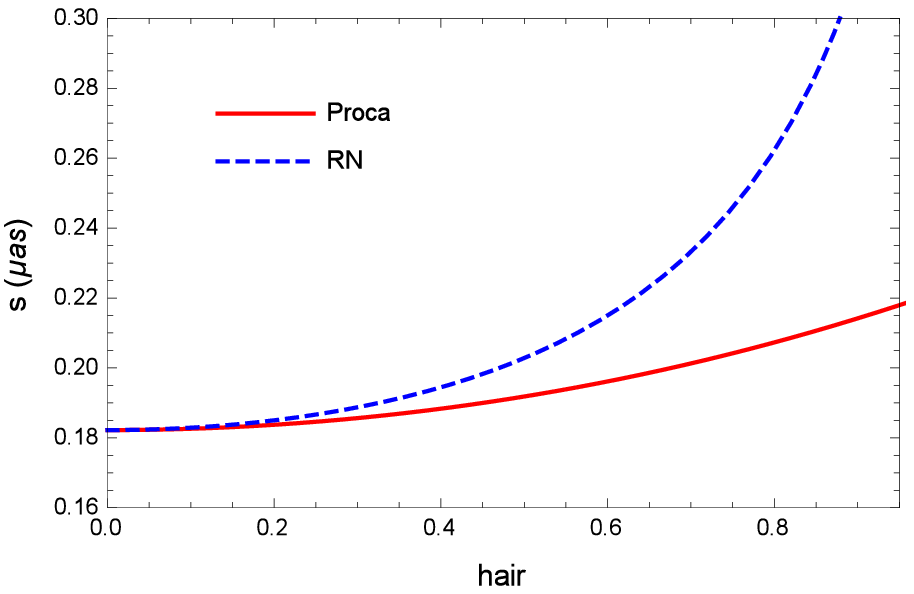}
		\end{minipage}
			\hfill
			\begin{minipage}[b]{0.45\textwidth}
				\includegraphics[width=\textwidth]{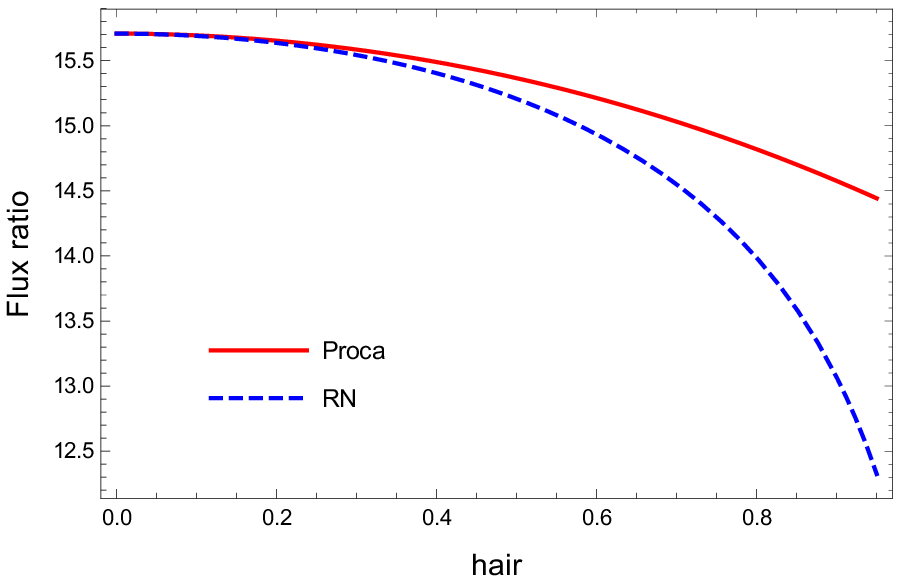}
			\end{minipage}
				\hfill
				\begin{minipage}[b]{0.45\textwidth}
					\includegraphics[width=\textwidth]{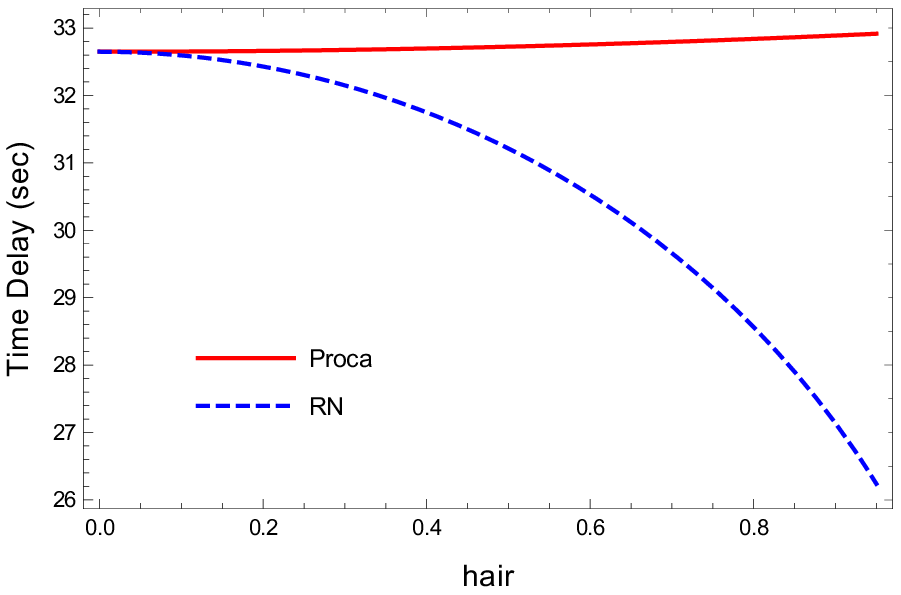}
				\end{minipage}
				\caption{Variation of different observables - (a) angular position of inner most image $\theta_{m}$, (top-left corner) (b) the angular difference between the outermost and inner images s , (top-right corner)  (c)  Flux ratio of innermost image with respect to the others, $\mathcal{R}$ (bottom-left corner) and (d) Time delay between first and second relativistic image that formed on the same side of the lens $\Delta T_{12}^{s}$ (bottom-right corner) as a function of the hair parameters. Here the parameter `$hair$' corresponds to Proca hair $P$ in the case of generalized Proca black holes  and charge hair $q$ in the case of Reissner-Nordstr\"{o}m black holes. The solid red lines indicates the behavior of the observables as function Proca hair $P$ for generalized Proca black holes whereas the blue dashed lines indices the variation of the observables as a function charge hair $q$ for Reissner-Nordstr\"{o}m black holes.}\label{observable}
	\end{figure}
%%%%%%%%%%%%%%%%%%%%%%%%%%%%%%%%%%%%%%%%%%%%%%%%	
		
		In Tab-\ref{tab:tab2}, `$hair$' corresponds to Proca hair $P$ in the case of generalized Proca black holes ( Eq. (\ref{intro5})) and charged hair $q$ in the case of Reissner-Nordstr\"{o}m (RN) black holes (Eq. (\ref{RN})). We also have plotted the observables as a function of hair parameter for these two black hole spacetime in Fig-\ref{observable}. From Table-\ref{tab:tab2}, we found out that in the case of generalized Proca black holes, angular position of the inner most image $\theta_{m}$ increases as the $hair$-parameter increases which is contrary to the RN case. This means that the size of the inner-most Einstein ring is bigger for the generalized Proca black hole space-time than those obtained from RN space-time for the same value of the hair parameter. The angular separation $s$ increases with the increase of the $hair$-parameter similar to case of RN black hole space-time while the Relative flux $\mathcal{R}$ decreases with the increase of the $hair$-parameter. From Tab-\ref{tab:tab2}, one can see that $\Delta T_{12}^{s}$ increases with the increase of $hair$- parameter for the case of generalized Proca black holes which is contrary to the RN case. Note that size of the inner-most Einstein ring  $\theta_{m}$(or, the time delay between first and second relativistic image $\Delta T_{12}^{s}$) is maximum for the case $q=0$ (Schwarzschild black hole) in static, spherically symmetric space-time predicted by General Relativity. So any value of  $\theta_{m}$ ( or $\Delta T_{12}^{s}$) greater than those predicted in Schwarzschild space-time implies the existence of Proca hair. Thus by measuring the size of the innermost Einstein ring (or the time delay delay between first and second relativistic image) in a static, spherically symmetric space-time, one can observationally verify the ``no-hair theorem"\cite{nohair}.

Now, in order to probe the Proca hair, one have to observationally measure both the position of innermost image $\theta_{m}$ and angular septation $s$. From the Table-\ref{tab:tab2}, we can see that the angular separation between the images is $\sim \mathcal{O}(10^{-1})$ $\mu$arcsec, which is too hard to detect with present technologies.

 Before doing further study with Proca black hole, we want to discuss an important issue regarding  the validity of our estimates for different astrophysical parameters. As mentioned in section \ref{sec2}, the solution (\ref{intro5}) for Proca black hole that we consider in our study, is not an exact solution of the Einstein equations, but an approximated analytical solution which agrees well with the full numerical solution upto order $\mathcal{O}(1/r^{3})$ . Although this is ok for regions away from the black hole horizon, but for near-horizon regions, analytical approximation may break down. To see how far it affects the numerical estimates of different observables, we consider solution upto order $\mathcal{O}(1/r^{4})$ ( next order) \cite{Heisenberg:2017hwb}, calculate different observables and study the percentage deviations from the corresponding values for solution upto order $\mathcal{O}(1/r^{3})$. The result is shown in Tab-\ref{table3}. As one can see, for most of the cases, the deviation is around $1\%$ or less, except for the parameter $s$ with high value of Proca hair, when the deviation is around $3\%$. Hence, as long as the errors in future observational estimates for these parameters are larger than these percentage deviations, our results are reliable.\\
%%%%%%%%%%%%%%%%%%%%%%%%%%%%%%%%%%%%%%%%%%%%%%%
\begin{table}[]
\centering
\def\arraystretch{.7}      	
\setlength{\tabcolsep}{3em}
\begin{tabular}{|l|l|l|l|l|} 
\hline
Proca hair & \multicolumn{4}{l|}{\begin{tabular}[c]{@{}l@{}}Percentage modification of the value when $\mathcal{O}(1/r^{4})$ is included in the metric for \\the observables \end{tabular}}  \\ 
\cline{2-5}
           & $\theta_{m}$  & $s$        & $\mathcal{R}$  & $\Delta T_{12}^{s}$                                                                                                                 \\ 
\hline
0.3        & 0.0550516     & 0.444458 & 0.111209       & 0.0550516                                                                                                                           \\
0.6        & 0.214403      & 1.70292  & 0.447742       & 0.214403                                                                                                                            \\
0.9        & 0.4623        & 3.57128  & 1.03114        & 0.4623                                                                                                                              \\
\hline
\end{tabular}
				\caption{ In Tab-\ref{tab:tab2}, numerical estimation of different observables related to strong lensing ($\theta_{m}$, $s$, $\mathcal{R}$, $\Delta T_{12}^{s}$) for metric (\ref{intro5}) have been presented where we have considered metric components upto order $\mathcal{O}(1/r^{3})$ only. This table shows the percentage modification of the observables when contribution from next leading order ($\mathcal{O}(1/r^{4})$) is taken into account. As one can see for most of the cases, the deviation is below $1\%$. }\label{table3}
\end{table} 
%%%%%%%%%%%%%%%%%%%%%%%%%%%%%%%%%%%%%%%%%%%%%%%%%%%%%%%%%%%%%%%%%%%%%%%%%%%%%%%%%%%%%%%%%%%%%%%%%%%%%%%%%%%%%%%%%%%%%%%%%%%%%%%%%%%%%%%%%%%%%%%%

\section{Lensing of Rotating Proca Black holes}

In the previous section, we have studied the bending of light ray trajectory in the presence of a static and spherically symmetric black hole. But several observations indicate that the super massive black hole in the center of our galaxy possesses angular momentum \citep{spinSgr}. So for observational perspective, it is important to consider the lensing effect for a rotating black hole. Moreover, the spacetime geometry is much more richer in this case. So from pure theoretical point of view, we can expect some interesting result will emerge when we consider strong lensing effect around a rotating black hole. Indeed, previous studies has showed that the caustic points are no longer aligned with the optical axis for the rotating black hole, but shifted in according to the rotation of the black hole and now they have a finite extension \cite{Bozza:2005tg}. In this present section, we will discuss  the gravitational lensing effect for a more general rotating black hole. First we calculate the metric for a rotating black hole in generalized Proca theories using Newman-Janis algorithm and then study the null trajectories in this spacetime. 
%%%%%%%%%%%%%%%%%%%%%%%%%%%%%%%%%%%%%%%%%%%%%%%%%%%%%%%%%%%%%%%%%%%%%%%%%%%%%%%%%%%%%%%%%%%%%%%%%%%%%%%%%%%%%%%%%%%%%%%%%%%%%%%%%%%%%%%%%%%%%%%%

\subsection{Null Geodesic equation and Photon Trajectory}
Applying Newman-Janis algorithm \cite{NJ1} to the metric.(\ref{intro5}) and retaining only terms up to the order $\mathcal{O}(\frac{P^{2}}{r^{3}})$, we found out the stationary, axisymmetric solution to Einstein's field equation which in Boyer-Lindquist coordinates $(t,r,\vartheta,\phi)$ \cite{BL} can be written as
%%%%%%%%%%%%%%%%%%%%%%%%%%%%%%%%%%%%%%%%%%%%%%%%
	 		\begin{eqnarray}{\label{101}}
	ds^{2}&=&-\left[1-\frac{2r}{\rho^{2}}-\frac{P^{2}}{6\rho^{2}r}\right]dt^{2}-\dfrac{4a \sin^{2}\vartheta}{\rho^{2}}\left[r+\frac{P^{2}}{8}+\frac{P^{2}}{6r}\right]dt d\phi +\dfrac{\rho^{2}}{\Delta}dr^{2}+\rho^{2}d\vartheta^{2}+ \nonumber\\
&&\left[r^{2}+a^{2}+\dfrac{2ra^{2}\sin^{2}\vartheta}{\rho^{2}}+\dfrac{ P^{2}a^{2}\sin^{2}\vartheta}{2\rho^{2}}\left(1+\frac{1}{r}\right)\right]\sin^{2}\vartheta d\phi^{2}
	 		\end{eqnarray}
%%%%%%%%%%%%%%%%%%%%%%%%%%%%%%%%%%%%%%%%%%%%%%%%
	 		where $a=L/M^{2}$, where  $L$ and $M$ denotes the angular momentum and mass of the black hole respectively. 
 The Proca field in this case is turned out to be
%%%%%%%%%%%%%%%%%%%%%%%%%%%%%%%%%%%%%%%%%%%%%%%%%%%%%%%%
%%%%%%%%%%%%%%%%%%%%%%%%%%%%%%%%%%%%%%%%%%%%%%%%%%%%%%%%
\begin{equation}\label{Proca}
\mathcal{A}_{\mu}=\left(\tilde{\mathcal{A}_{0}},-\frac{\tilde{\mathcal{A}}_{0}\rho^{2}}{\Delta}\frac{1}{\sqrt{\tilde{A}\tilde{B}}}+\left(\tilde{\mathcal{A}}_{0}+\tilde{\mathcal{A}}_{1}\sqrt{\frac{\tilde{B}}{\tilde{A}}}\right)\left(1-\frac{a^{2}\sin^{2}\theta}{\Delta}\right),0,\tilde{\mathcal{A}}_{1}\sqrt{\frac{\tilde{B}}{\tilde{A}}}a\sin^{2}\theta\right)
\end{equation}
%%%%%%%%%%%%%%%%%%%%%%%%%%%%%%%%%%%%%%%%%%%%%%%%%%%%%%%%%%
%%%%%%%%%%%%%%%%%%%%%%%%%%%%%%%%%%%%%%%%%%%%%%%%%%%%%%%%%
where ``tilde'' denotes the components after complexification. We have checked that eqn (23) and (24) together satisfy the Einstein's equation for an axisymmetric metric for a rotating Proca black hole upto order  $\mathcal{O}(1/r^{4})$. As our original non-rotating black hole solution is valid upto order $\mathcal{O}(1/r^{3})$, we can safely take the metric given by eqn (23) for a rotating Proca black hole for further study.

		We can also identify the quantity $a$  in eqn (23) as the specific angular momentum of the black hole. The functions $\rho$ and $\Delta$ is given by
%%%%%%%%%%%%%%%%%%%%%%%%%%%%%%%%%%%%%%%%%%%%%%%
	 		\begin{align}{\label{102}}
	 		\rho^{2}&=r^{2}+a^{2}\cos^{2}\vartheta\\\Delta&=r^{2}+a^{2}-2r-\frac{P^{2}}{2}\left(1+\frac{1}{r}\right)
	 		\end{align}
%%%%%%%%%%%%%%%%%%%%%%%%%%%%%%%%%%%%%%%%%%%%%%%%
	 		Here also, we have set $r=r/M$ and $P=P/M_{Pl}$. In the limit of vanishing Proca hair $i.e.$ $P\to 0$, the solution coincides with the Kerr black hole. Horizon of the black hole is a surface where $\Delta=0$ and outer horizon is determined by the largest possible solution of the equation, which in this case turns out to be
%%%%%%%%%%%%%%%%%%%%%%%%%%%%%%%%%%%%%%%%%%%%%%		
	 		\begin{equation}{\label{103}}
	 		r_{H}=\dfrac{2}{3}-\dfrac{-16+12a^{2}-6P^{2}}{3\sqrt[3]{4}\sqrt[3]{4 \sqrt{K}-144 a^2+180 P^2+128}}+\dfrac{\sqrt[3]{4 \sqrt{K}-144 a^2+180 P^2+128}}{6\sqrt[3]{4}}
	 		\end{equation}
%%%%%%%%%%%%%%%%%%%%%%%%%%%%%%%%%%%%%%%%%%%%%%	 		
	 		where the function $K=2 \left(6 a^2-3 P^2-8\right)^3+\left(-36 a^2+45 P^2+32\right)^2$. It is easy to see in the limit $a,P\to 0$, radius of the horizon is turns out to be $r_{H}=2$ as expected for Schwarzschild case.
	 		
	 		Null geodesics equations can be obtained by using Hamilton-Jacobi Equation \cite{chandra}. For the metric.(\ref{101}), the relevent geodesic equations are given by
%%%%%%%%%%%%%%%%%%%%%%%%%%%%%%%%%%%%%%%%%%%%%		
	 		\begin{align}
	 		\rho^{2}\dot{r}&=\sqrt{{R}(r)}{\label{104}}\\\rho^{2}\dot{\vartheta}&=\sqrt{\Theta(\vartheta)}{\label{105}}\\\rho^{2}\dot{\phi}&=\frac{a}{\Delta}\left(r^{2}+a^{2}-aJ\right)+\dfrac{P^{2}a}{\Delta}\left(\frac{1}{2}+\frac{1}{3r}\right)+\left(J\sin^{-2}\vartheta-a\right){\label{106}}
	 		\end{align}
%%%%%%%%%%%%%%%%%%%%%%%%%%%%%%%%%%%%%%%%%%%%%%	 		
where
%%%%%%%%%%%%%%%%%%%%%%%%%%%%%%%%%%%%%%%%%%%
	 		\begin{align}
	 		{R}(r)&=\left(r^{2}+a^{2}-aJ\right)^{2}-P^{2}a\left(\frac{1}{2}+\frac{1}{3r}\right)-\Delta\left(Q+(J-a)^{2}\right){\label{107}}\\\Theta(\vartheta)&=Q-\left(J^{2}\sin^{-2}\vartheta-a^{2}\right)\cos^{2}\vartheta{\label{108}}
	 		\end{align}
%%%%%%%%%%%%%%%%%%%%%%%%%%%%%%%%%%%%%%%%%%%%	 		
	 		In Eq. (\ref{104})-(\ref{106}), the dot indicates derivative w.r.t. some affine parameter $\lambda$. In Eq. (\ref{107}) and Eq. (\ref{108}), $Q$ denotes a constant of separation called Carter constant, $J$ is the angular momentum of the photon with respect to the axis of the black hole. In our analysis, we have set $p_{t}=-E=-1$ by a suitable choice of affine parameter.

      Since we are interested in studying the photon trajectory in an isolated black hole spacetime, we can ignore the effect of other celestial bodies on the photon trajectory and can approximate the spacetime at a large distance from the black hole as flat Minkowski spacetime. We will assume that both the source and observer are situated at a large distance from the black hole. Now we can formulate the lensing problem as follows: the initial photon trajectory starts off as a straight line. If there were no black hole, then it would continue to follow this straight line trajectory. But because of the presence of the black hole, its path gets deviated from this initial trajectory near the black hole. Finally, it approaches the observer along this deviated path. In this scenario, we can relate the constant of motion $Q$ and $J$ in terms of a set of geometric quantities  $(\psi_{R},u,h)$ \cite{Bozza:2002af}. Here the inclination angle $\psi_{R}$ denotes the angle between the initial photon trajectory and the equatorial plane. The projected impact parameter $u$ describes the minimum distance of projected photon trajectory on the equatorial plane from the origin if there were no black hole and lastly, the height of the light ray trajectory at $u$ from the equatorial plane is denoted by $h$.
	 		
	 		Let $(\alpha,\beta)$ denotes the celestial coordinate of the image as seen by an observer sitting on $(r_{R},\vartheta_{R})$  in Boyer-Lindquist coordinate. The coordinate $\alpha$ and $\beta$ represents the apparent perpendicular distance of the image from the axis of symmetry and its
	 		projection on the equatorial plane respectivily \cite{chandra}. Taking into consideration that the observer is situated far away from the black hole and using Eq. (\ref{104} -\ref{108}), we can express $\alpha$ and $\beta$ as \cite{chandra, Gyulchev:2006zg}
%%%%%%%%%%%%%%%%%%%%%%%%%%%%%%%%%%%%%%%%%%%	 		
	 		\begin{align}
	 		\alpha&=-r_{R}^{2}\sin\vartheta_{R}\dfrac{d\phi}{dr}\Biggm|_{r_{R}\to \infty}=\frac{J}{\sin\vartheta_{R}}\label{Sh1}\\\beta&=r_{R}^{2}\dfrac{d\vartheta}{dr}\Biggm|_{r_{R}\to \infty}=h\sin\vartheta_{R}\label{Sh2}
	 		\end{align}
%%%%%%%%%%%%%%%%%%%%%%%%%%%%%%%%%%%%%%%%%%%%	 		
	 		Taking the asymptotic limit $\vartheta_{R}=\pi/2-\psi_{R}$ and $\alpha=u$, we can finally express the constants of motion in terms of geometric parameters of the incoming ray as
%%%%%%%%%%%%%%%%%%%%%%%%%%%%%%%%%%%%%%%%%%%	 		
	 		\begin{align}
	 		J&\approx u \cos\psi_{R}{\label{109}}\\Q&\approx h^{2}\cos^{2}\psi_{R}+(u^{2}-a^{2})\sin^{2}\psi_{R}{\label{110}}
	 			 		\end{align}
%%%%%%%%%%%%%%%%%%%%%%%%%%%%%%%%%%%%%%%%%%%%%%%%%%%%%%%%%%%%%%%%%%%%%%%%%%%%%%%%%%%%%%%%%%%%%%%%%%%%%%%%%%%%%%%%%%%%%%%%%%%%%%%%%%%%%%%%%%%%%%%%%%%%%%%%%

\section{Black Hole Shadow analysis}

	 	In this section, we will describe the shadow of rotating black hole. For non-rotating black hole, it is just a black circular disc in the observable sky with a radius that corresponds to the position of the photon sphere. As we will see, things goes a bit interesting in the case of rotating black hole. 
	 	
	 	We will use the celestial co-ordinates $(\alpha,\beta)$ given in Eq. (\ref{Sh1})-(\ref{Sh2}) to give a description of the shadow. For simplicity let assume that the observer is sitting on the equatorial plane. Then using Eq. (\ref{Sh1})-(\ref{110}), it is easy to check that photons reaching from an generic point $(\alpha/r_{R},\beta/r_{R})$ can be characteristic by $J = -\alpha$ and $Q=\beta^{^2}$. In our calculation we have considered that positive angular momentum $J$ corresponds to counterclockwise winding of the light rays as seen from above. So when $a>0$, the photons rotates in the same direction as the black hole (prograde/direct photons) while they rotate in the opposite direction for  $a<0$ case (retrograde photons). One can visualize of the shape of black hole shadow by plotting $\beta$ vs $\alpha$. 
%%%%%%%%%%%%%%%%%%%%%%%%%%%%%%%%%%%%%%%%%%%%%	 	
	 		\begin{figure}
	 			\centering 
	 			\begin{minipage}[b]{0.45\textwidth}
	 				\includegraphics[width=\textwidth]{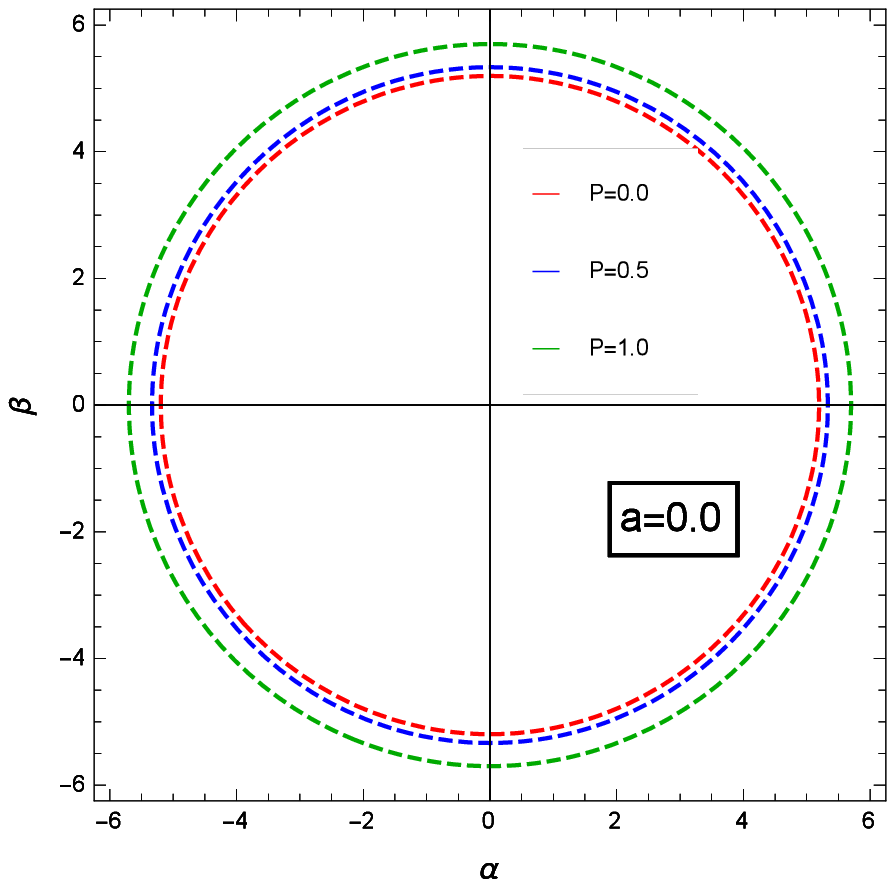}
	 			\end{minipage}
	 			\hfill
	 			\begin{minipage}[b]{0.45\textwidth}
	 				\includegraphics[width=\textwidth]{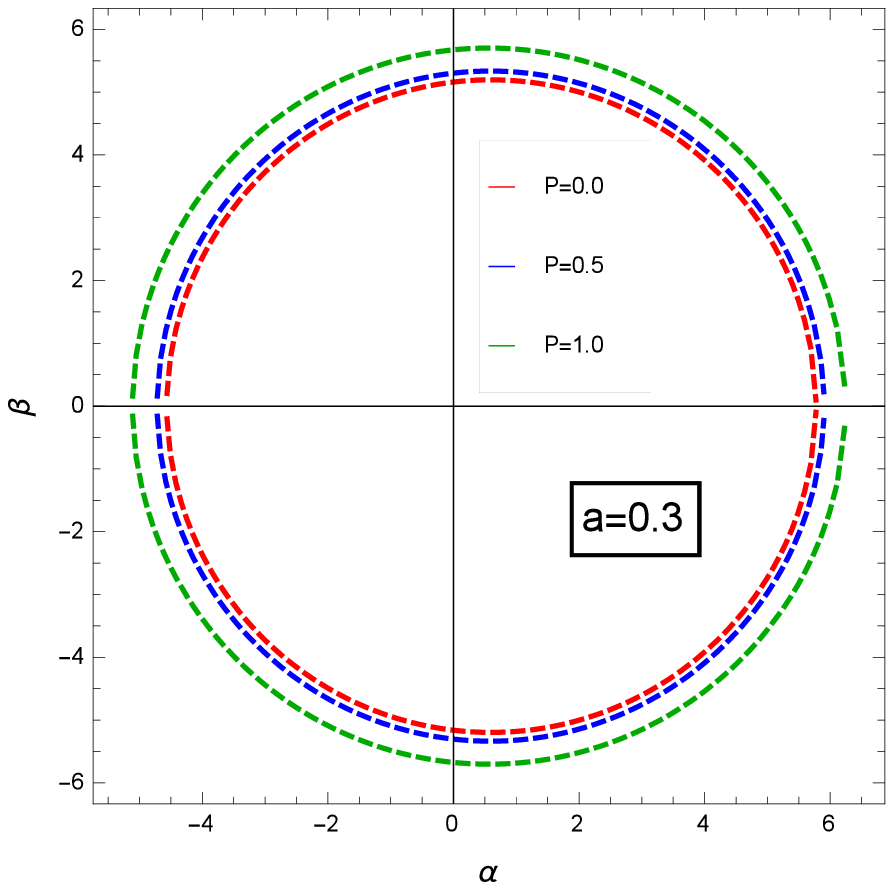}
	 			\end{minipage}
	 			\hfill
	 			\begin{minipage}[b]{0.45\textwidth}
	 				\includegraphics[width=\textwidth]{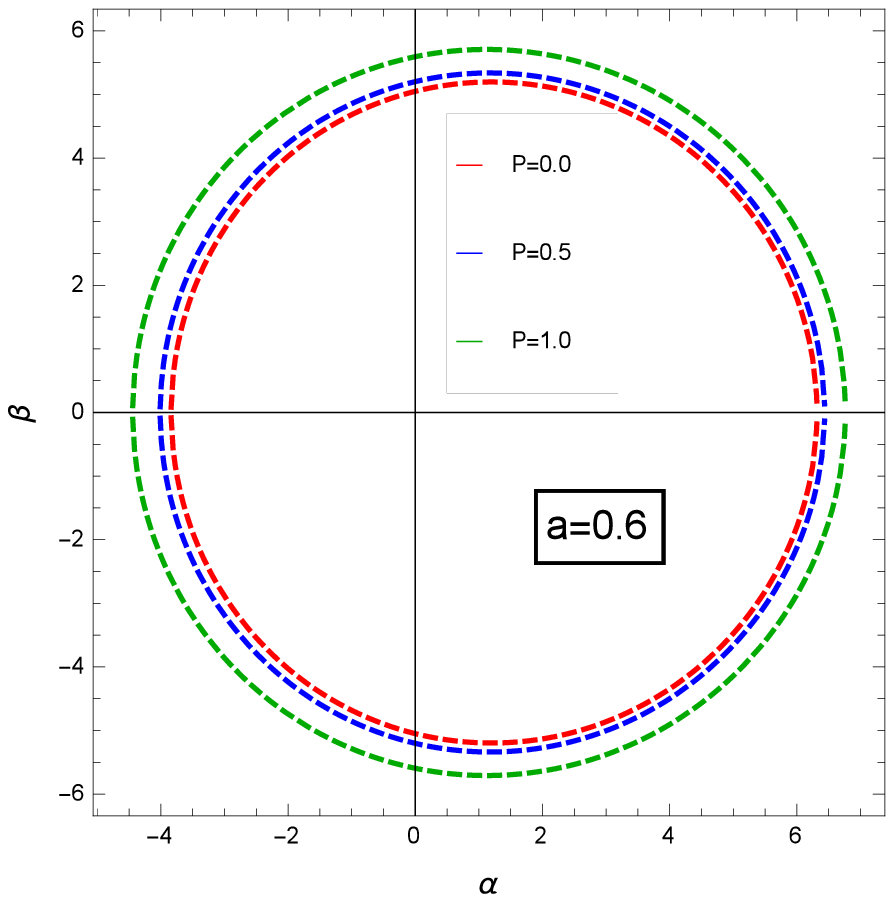}
	 			\end{minipage}
	 			\hfill
	 			\begin{minipage}[b]{0.45\textwidth}
	 				\includegraphics[width=\textwidth]{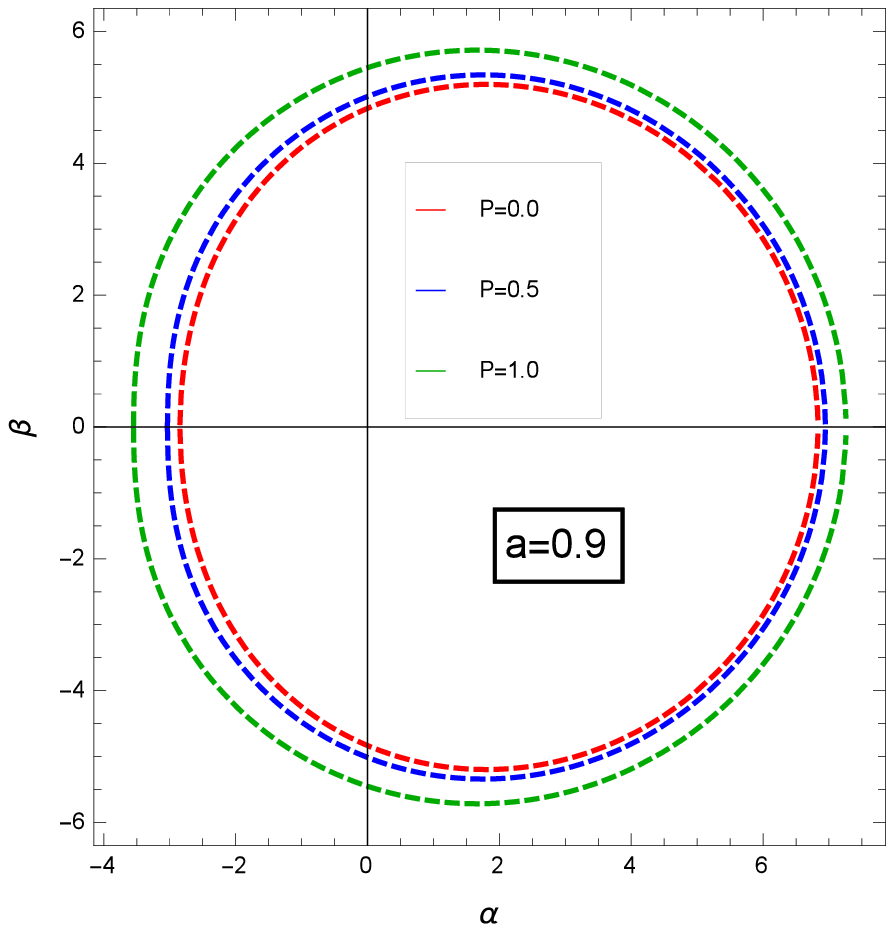}
	 			\end{minipage}
	 			\caption{Shadow casted by rotating black hole in generalized Proca theories given by the metric.(\ref{101}) for different values of $a$ and $P$ as seen by observer in equatorial plane. The shadow region is corresponds to the inside of each dashed curve. The case $a=0$ defines shadow for non-rotating black hole. An increase of $a$ causes the deformation of the black hole shadow.   }\label{Shgola}
	 		\end{figure}
%%%%%%%%%%%%%%%%%%%%%%%%%%%%%%%%%%%%%%%%%%%%%	 		
In Fig-\ref{Shgola}, we have plotted the shadows casted by a black hole described by metric.(\ref{101}) for different values of $a$ and $P$. For the non-rotating case ($a=0$), the shadow of the black hole is just a circular disc. When the spin of the black hole is non-vanishing, the shadow shape gets slightly distorted and gets displaced to the right. Physically which means that the prograde photons (photons coming from the left side as seen by the observer) are allowed to get closer to the black hole while the retrograde photons (photons coming from the right side as seen by the observer) are kept even further.

\begin{figure}
\includegraphics[width=.50\textwidth]{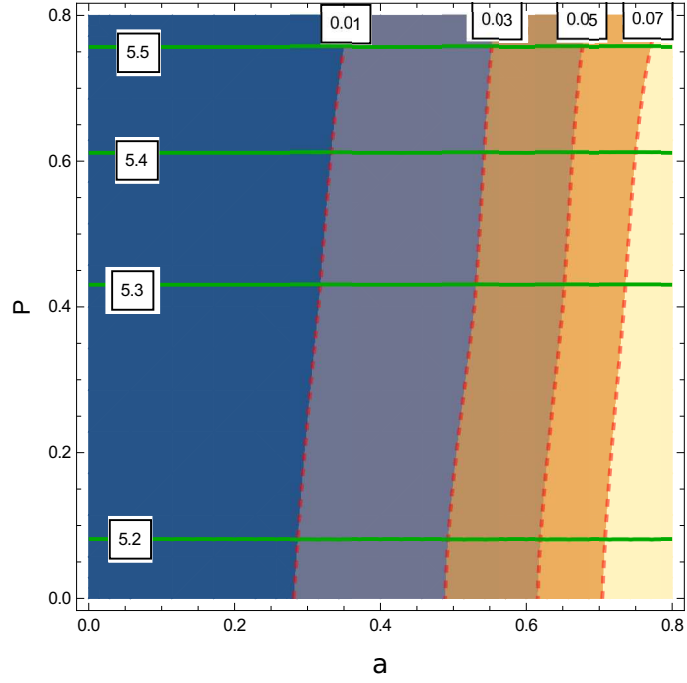}
\caption{The contour plot of constant $R_{s}$ (green solid lines)and $\delta_{s}$ (red dashed lines) curves in the $(a,P)$ plane have been presented. Intersection of the curves corresponding to constant $R_{s}$ and $\delta_{s}$ obtained from observation gives value of $a$ and $P$ of the black hole.}
\label{ap}
\end{figure}

	 	Following Ref. \cite{Hioki:2009na}, we define the observables for black hole shadow as the radius $R_{s}$ of a reference circle and the distortion parameter $\delta_{s}$. We will consider a reference circle that passes through three points of the shadow: the top $(\alpha_{t},\beta_{t})$ and the bottom $(\alpha_{b},\beta_{b})$ and a point corresponds to the unstable retrograde circular orbit $(\alpha_{r},0)$. The distortion parameter $\delta_{s}$ is the ratio of the difference between the endpoint of the circle $(\bar{\alpha}_{p},0)$ and the point corresponding to the prograde circular orbit $(\alpha_{p},0)$ (both of them at the opposite side of the point $(\alpha_{r},0)$) to radius of the reference circle \cite{Amarilla:2011fx}. Typically $R_{s}$ gives the approximate size of the black hole shadow, while $\delta_{s}$ is a measure of its deformation w.r.t. the reference circle. For an equatorial observer, the observables takes the form
%%%%%%%%%%%%%%%%%%%%%%%%%%%%%%%%%%%%%%%%%%%%	 	
	 		\begin{align}
	 	R_{s}&=\dfrac{(\alpha_{t}-\alpha_{r})^{2}+\beta_{t}^{2}}{2|\alpha_{t}-\alpha_{r}|}{\label{Sh3}}\\\delta_{s}&=\dfrac{\bar{\alpha}_{p}-\alpha_{p}}{R_{s}}{\label{Sh4}}
	 		\end{align}
%%%%%%%%%%%%%%%%%%%%%%%%%%%%%%%%%%%%%%%%%%%%%	 		
	 		By measuring this two observables, one can predict the black hole parameters very accurately.
	 		A simple way to extract the information about parameters $a$ and $P$ is to plot the contour curves of constant $R_{s}$ and $\delta_{s}$ in the $(a,P)$ plane \cite{Amarilla:2013sj}. The points in that plane where they intersect give the value of corresponding $a$ and $P$.  In Fig--\ref{ap}, we show the contour plot of $R_{s}$ and $\delta_{s}$ in the $(a,P)$ plane. As stated earlier, if we can obtain values of $R_{s}$ and  $\delta_{s}$ very accurately from the observations, the point where the associated contours  intersect, gives the corresponding values of $a$ and $P$.
\section{Gravitational lensing by a rotating black hole in strong field limit  }
In this section, we briefly review the the main concepts and the observables related to strong lensing in a stationary, axisymmetric space-time following the methods developed by Bozza \cite{Bozza:2002af}. Throughout our discussion, we have considered that both the source and observer are situated very far away from the black hole. For sake of simplicity, we restrict our attention to the trajectories that are very close to equatorial plane. Advantage of considering such scenario is that the angular position of the images can still be described by those obtained in the equatorial plane but now one can understand the problem in some deeper level as one can calculate the magnification of the images from two dimensional lens equation. \\
We formulate the lensing problem as follows : we consider that the observer and the source are situated at a height $h_{R}$ and $h_{S}$ from the equatorial plane respectively. A photon with impact parameter $u$ incoming from the source situated at $r_{S}$, approaches the black hole at a minimum distance $r_{0}$ and then deviates away from it. An observer at $r_{R}$ receives the photon. Now we want to find the angular position and the magnification of the images.
In order to do so, we first restrict our attention to the light rays on the equatorial plane by setting $\vartheta=\pi/2$ or equivalently by taking $\psi=\pi/2-\vartheta=0$ and $h=0$. Substituting these conditions on Eq. (\ref{101}), we get reduced metric of the form
\begin{equation}{\label{111}}
ds^{2}=-A(r)dt^{2}+B(r)dr^{2}+C(r)d\phi^{2}-D(r)dt d\phi
\end{equation}
 In order to study the photon trajectory in a stationary space-time, we assume that the equation \cite{Gyulchev:2006zg}
\begin{equation}\label{123}
(A_{0}C'_{0}-A_{0}'C_{0})^{2}=(A_{0}'D_{0}-A_{0}D'_{0})(C_{0}'D_{0}-C_{0} D'_{0})
\end{equation}
admits at least one positive solution and largest positive root of the equation is defined as the radius of photon sphere, $r_{m}$.	Here subscript `$0$' implies functions evaluated at closest approach distance $r_{0}$. Note that, when we put $D_{0}=0$, this equation coincides with the condition for photon sphere in static case given by Eq. (\ref{2}). In the strong field limit, we consider only those photons whose closest approach distance $r_{0}$ is very near to $r_{m}$ and hence the deflection angle $\alpha$ can be expanded around the photon sphere, $r_{m}$ or equivalently minimum impact parameter $u_{m}$. When the closest approach distance $r$ is greater than $r_{m}$, it just simply gets deflected (it may complete several loops around the black hole before reaching the observer). When it reaches a critical value $r_{0}=r_{m}$ (or $u=u_{m}$), $\alpha$ diverges and the photon gets captured. Using the same method as in the case static case, we can express total deflection as follows \cite{Bozza:2002af}
\begin{equation}\label{124}
\alpha_{f}(\theta)=-\bar{a}_{rot}\log\left(\dfrac{\theta}{\theta_{m}}-1\right)+\bar{b}_{rot}
 \end{equation}
 where $u_{m}$ is the impact parameter evaluated at $r_{m}$. The parameters $\bar{a}_{rot}$, $\bar{b}_{rot}$ are the Strong field coefficient for rotating metric in the equatorial plane (for explicit expression of  $\bar{a}_{rot}$, $\bar{b}_{rot}$, see Eq. (34-35) of Ref \cite{Bozza:2002af} ). $\theta $ denotes the angular position of the image. From Eq. (\ref{124}), we can see that the deflection angle diverges at $\theta=\theta_{m}=u_{m}/r_{R}$. It represents the position of the innermost image. Once the deflection angle is known, we can obtain the angular position of different images using the lens equation (\ref{8}). In the simplest situation, one can express the angular position of the n-th order image as \cite{Bozza:2002af}
     \begin{equation}\label{129} 
    \theta_{n}=\theta_{n}^{0}\left[1-\dfrac{u_{m}e_{n}}{\bar{a}_{rot}}\left(\dfrac{r_{R}+r_{S}}{r_{R}r_{S}}\right)\right]
    \end{equation}
    where, 
    \begin{eqnarray}
& &{\theta}_{n}^{0}=\theta_{m}(1+e_{n})\,,\qquad 
{e}_{n}=\exp\left[{\dfrac{\bar{b}_{rot}+\gamma-2n\pi}{\bar{a}_{rot}}}\right]\,. \nonumber 
\end{eqnarray}
Now we turn our attention to the trajectories that are very close to the equatorial plane. i. e. those trajectories have very small value of declination angle $\psi=\pi/2-\vartheta$. With the help of this condition and assuming the height of light ray trajectory from equatorial plane $h$ is small compared to the projected impact parameter $u$, it is easy to show that the inclination angle $\psi_{R}\approx h/u$. Then the constants of motion are given in  Eq. (\ref{109})-(\ref{110}) can be written as
  \begin{eqnarray}\label{130}
 & &{J}\approx u\,,\qquad 
{Q}\approx h^{2}+\bar{u}^{2}\psi_{R}^{2}\,. 
  \end{eqnarray}

  where $\bar{u}=\sqrt{u^{2}-a^{2}}$. Moreover, we expect the declination angle $\psi$ to remain small (of the order of $\psi_{R}$) during the motion. Small declination condition for the photon trajectory readily implies that $(h_{R},h_{S})\ll u\ll (r_{R},r_{S})$. If we neglect the higher order terms, then the polar lens equation can be written as \cite{Bozza:2002af}
	 \begin{equation}\label{140}
h_{S}=h_{R}\left[\frac{r_{R}}{\bar{u}}\sin\bar{\phi}_{f}-\cos\bar{\phi}_{f}\right]-\psi_{R}\left[(r_{R}+r_{S})\cos\bar{\phi}_{f}-\dfrac{r_{R}r_{S}}{\bar{u}}\sin\bar{\phi}_{f}\right]
	 \end{equation}
	 where 
	 \begin{equation}\label{137}
	  \bar{\phi}_{f}(\theta)=-\hat{a}_{rot}\ln\left(\dfrac{\theta}{\theta_{m}}-1\right)+\hat{b}_{rot}
	 \end{equation}
	 Here, $\hat{a}_{rot}$ and $\hat{b}_{rot}$ denotes two numerical parameters (For more details see Eq. (52-53) of Ref \cite{Bozza:2002af}). Eq. (\ref{140}) along with Eq. (\ref{8}) represents the two dimensional lensing equation. Using these two equation one can find we get the magnification of the $n$-th image as
   \begin{equation}\label{147}
  \mu_{n}=\dfrac{(r_{R}+r_{S})^{2}}{(r_{R}r_{S})}\left(\frac{\bar{\mu}(a)}{\mathcal{K}(\gamma)}\right)
   \end{equation} 
   where
    \begin{eqnarray}\label{148}
 & &{\bar{\mu}}(a)=\dfrac{\bar{u}_{m}(a)u_{m}(a)e_{\gamma}}{\hat{a}_{rot}(a)}\,,\qquad 
{e}_{\gamma}=\exp{(\dfrac{\hat{b}_{rot}+\gamma}{\hat{a}_{rot}})} \,,\qquad 
{\mathcal{K}}(\gamma)=r_{R}r_{S}\sin{\bar{\phi}_{f,n}}-\bar{u}_{m}(r_{S}+r_{R})\cos{\bar{\phi}_{f,n}}\,.
  \end{eqnarray}
      where ${\bar{\phi}}_{f,n}$ is the phase of the $n$-th order image given by the Eq. (\ref{137}) with $\theta_{n}$ is the solution of Eq. (\ref{129}). Note that $\mu_{n}$ diverges when $\mathcal{K}(\gamma)$ vanishes. This condition gives the position of the caustic points which formally defined as the positions of source for which one gets infinite magnification of the  images.     	
      \section{Time Delay between Different Images in Stationary space-time} 
   
     In this section we will extend our study of time delay effect in a rotating black hole space-time. As stated earlier, formation of different image is a result of photons following different trajectories, time taken by different photons is not the same and hence there will be a time delay between them. Bozza \cite{Bozza:2003cp} solved this problem using the same method used to find deflection angle and showed that the leading term in time delay between $m$-th and $n$-th order image can be expressed as \cite{Bozza:2003cp}
 \begin{equation}\label{Td6}
\Delta T_{mn}^{s}=2\pi (n-m)\frac{\widetilde{a}_{rot}(a)}{\bar{a}_{rot}(a)}
 \end{equation}
 When both the images are formed on the same side of the black hole. We can see that the time delay for direct photons ($a>0$) and retrograde photons ($a<0$) will be different.

 When images are formed on the opposite side of the lens, then time delay between $m$-th and $n$-th order image can be expressed as \cite{Bozza:2003cp}
 \begin{equation}\label{Td7}
 \Delta T_{mn}^{o}=\frac{\widetilde{a}_{rot}(a)}{\bar{a}_{rot}(a)}\left(2\pi n+\gamma-\bar{b}_{rot}(a)\right)+\widetilde{b}_{rot}(a)-\frac{\widetilde{a}_{rot}(-a)}{\bar{a}_{rot}(-a)}\left(2\pi n-\gamma-\bar{b}_{rot}(-a)\right)-\widetilde{b}_{rot}(-a)
 \end{equation}
 This extra contribution comes due to the fact that the co-efficient $\bar{b}_{rot}$ and $\widetilde{b}_{rot}$ is not same for direct and retrograde photons in the stationary case. The functional form of $\widetilde{a}_{rot}(a)$ and $\widetilde{b}_{rot}(a)$   is given in Eq. (35)-(36) in Ref \cite{Bozza:2003cp}.
      		\begin{table}
\centering
      	      	\def\arraystretch{1}
      	      	\setlength{\tabcolsep}{1em}
\begin{tabular}{|l|l|l|l|l|l|l|l|l|l|} 
\hline
\multirow{2}{*}{ $a$ } & \multirow{2}{*}{$hair$} & \multicolumn{2}{l|}{$\theta_{m}$ in $\mu$as } & \multicolumn{2}{l|}{$s$ in $\mu$as } & \multicolumn{2}{l|}{$\mathcal{R}$} & \multicolumn{2}{l|}{Time Delay $\Delta T^{s}_{12}$ }  \\ 
\cline{3-10}
                       &                    & Proca BH   & KN BH                                 & Proca BH    & KN BH                      & Proca BH   & KN BH            & Proca BH   & KN BH                                          \\ 
\hline
\multirow{4}{*}{0.0}   & 0.0                & 19.0033 & 19.0033                             & 0.0237825 & 0.0237825                & 15.708  & 15.708         & 32.6484 & 32.6484                                     \\
                       & 0.2                & 19.0103 & 18.8756                             & 0.0241503 & 0.0243711                & 15.6536 & 15.6367        & 32.6605 & 32.429                                      \\
                       & 0.4                & 19.0313 & 18.4796                             & 0.0252867 & 0.0263996                & 15.4896 & 15.4039        & 32.6966 & 31.7488                                     \\
                       & 0.6                & 19.066  & 17.7691                             & 0.0272943 & 0.0310206                & 15.2132 & 14.934         & 32.7563 & 30.5281                                     \\ 
\hline
\multirow{4}{*}{0.1}   & 0 .0               & 18.2607 & 18.2607                             & 0.0295151 & 0.0295151                & 15.708  & 15.708         & 31.3726 & 31.3726                                     \\
                       & 0.2                & 18.2666 & 18.1332                             & 0.0300417 & 0.0302463                & 15.6413 & 15.6371        & 31.3829 & 31.1537                                     \\
                       & 0.4                & 18.2845 & 17.7382                             & 0.031675  & 0.0327568                & 15.4398 & 15.4058        & 31.4135 & 30.475                                      \\
                       & 0.6                & 18.314  & 17.0297                             & 0.0345836 & 0.0384271                & 14.0986 & 14.9395        & 31.4642 & 29.2577                                     \\ 
\hline
\multirow{4}{*}{0.2}   & 0.0                & 17.4932 & 17.4932                             & 0.0371839 & 0.0371839                & 15.708  & 15.708         & 30.054  & 30.054                                      \\
                       & 0.2                & 17.4978 & 17.3667                             & 0.0379561 & 0.0380791                & 15.6247 & 15.6388        & 30.0619 & 29.8367                                     \\
                       & 0.4                & 17.5115 & 16.9749                             & 0.0403633 & 0.0411289                & 15.3721 & 15.4133        & 30.0855 & 29.1635                                     \\
                       & 0.6                & 17.5343 & 16.2738                             & 0.0446884 & 0.0478844                & 15.9415 & 14.9618        & 30.1247 & 27.9591                                     \\ 
\hline
\multirow{4}{*}{0.3}   & 0.0                & 16.6958 & 16.6958                             & 0.0476924 & 0.0476924                & 15.708  & 15.708         & 28.684  & 28.684                                      \\
                       & 0.2                & 16.6986 & 16.5713                             & 0.048857  & 0.048752                 & 15.6018 & 15.6427        & 28.6889 & 28.4701                                     \\
                       & 0.4                & 16.707  & 16.1864                             & 0.0525067 & 0.0523042                & 15.2765 & 15.4316        & 28.7033 & 27.8089                                     \\
                       & 0.6                & 16.7209 & 15.5015                             & 0.0591226 & 0.0598653                & 14.7165 & 15.0155        & 28.7272 & 26.6322                                     \\
\hline
\end{tabular}
      	      	\caption{Numerical estimations of the observables related to strong lensing ($\theta_{m}$, $s$, $\mathcal{R}$, $\Delta T_{12}^{s}$) by a rotating black hole have been presented. A comparison between the values of the observables obtained from rotating black holes in generalized Proca theories ( Proca BH) to those obtained from \KN black hole ( KN BH) have also been presented. Here the parameter `$hair$' corresponds to Proca hair $P$ in the case of rotating Proca BH and charged hair $q$ in the case of \KN black holes. Note that, $hair=0$ case corresponds to Kerr black hole. Here the SMBH at the center of our galaxy is taken as the lens. We have assumed that the lensing system is highly aligned and both the source and observer are situated at infinity. The time delay between the first and second relativistic image $\Delta T_{12}^{s}$ have been calculated under the assumption that both of these images are formed on the same side of the lens.}
      	      	\label{Tab3}
\end{table}      	
         	\begin{figure}
         		\centering
         			\begin{minipage}[b]{0.45\textwidth}
         		\includegraphics[width=\textwidth]{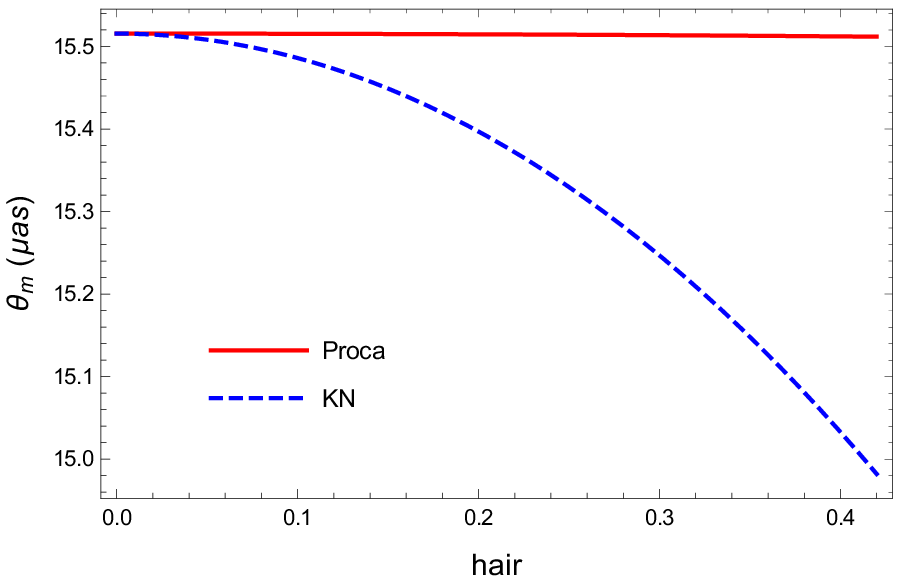}
         	\end{minipage}
         	\hfill
         		\begin{minipage}[b]{0.45\textwidth}
         		\includegraphics[width=\textwidth]{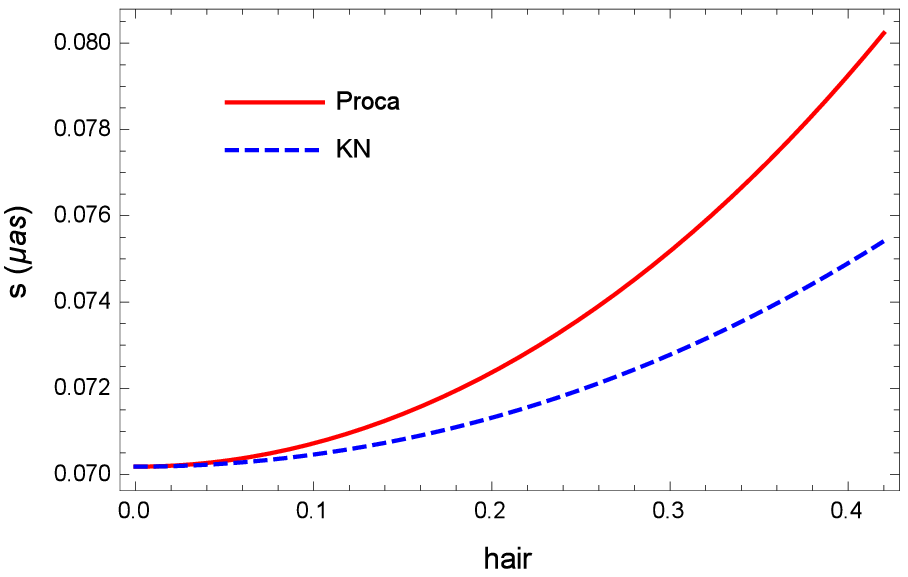}
         	\end{minipage}
         	\hfill
     	 	 	\begin{minipage}[b]{0.45\textwidth}
         		\includegraphics[width=\textwidth]{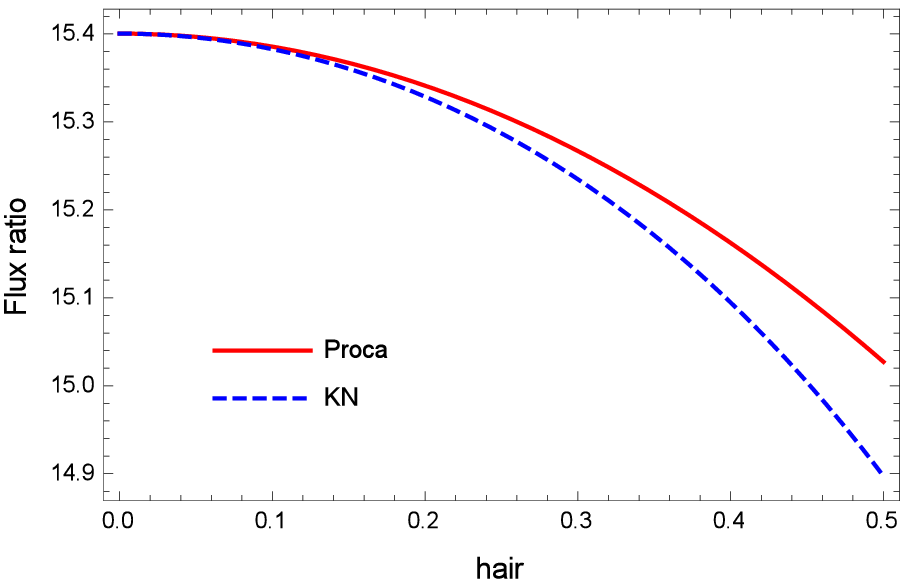}
         	\end{minipage}
         	\hfill
         	\begin{minipage}[b]{0.45\textwidth}
         		\includegraphics[width=\textwidth]{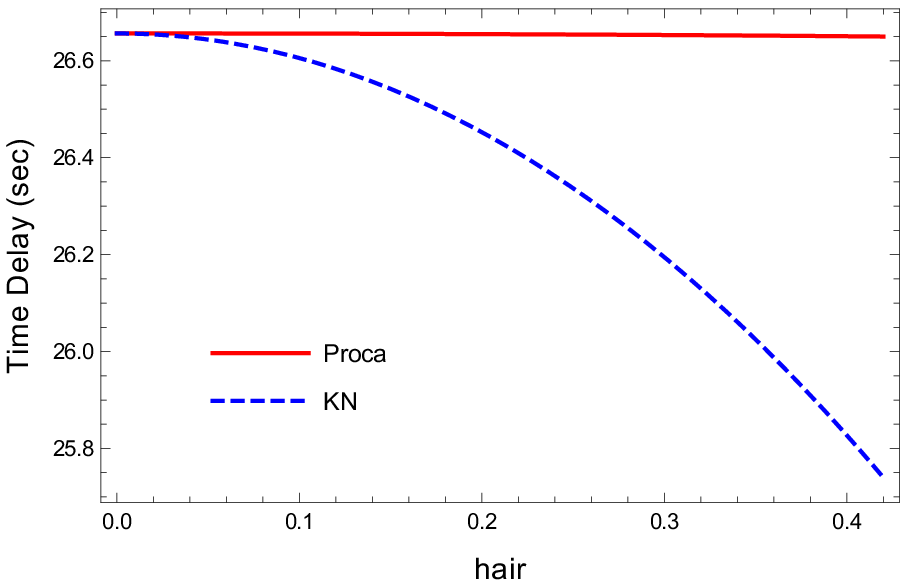}
         	\end{minipage}
         	\caption{Variation of different observables - (a) angular position of inner most image $\theta_{m}$, (top-left corner) (b) the angular difference between the outermost and inner images s , (top-right corner)  (c)  Flux ratio of innermost image with respect to the others, $\mathcal{R}$ (bottom-left corner) and (d) Time delay between first and second relativistic image $\Delta T^{s}_{12}$ when both of these images are formed on the same side of the lens (bottom-right corner) as a function of the hair parameters have been presented. Here the parameter `$hair$' corresponds to Proca hair $P$ in the case of rotating black hole in generalized Proca black holes  and charge hair $q$ in the case of \KN black holes. The solid red lines indicates the behavior of the observables as function Proca hair $P$ for generalized Proca black holes whereas the blue dashed lines indices the variation of the observables as a function charge hair $q$ for \KN black holes. Here we have assumed that the lensing system is highly aligned and both the source and observer are situated at infinity. We have taken that the  value of the black hole spin is taken as $a=0.44$, which is the current estimated value of spin of Sgr A* \cite{spinSgr}. }
         	\label{observablesspin}
         \end{figure}
\section{Numerical Estimation of different Observational Parameters}
In this section, we present the numerical estimation of different observational parameters related to strong lensing for stationary, axiasymmetric spacetime considering the SMBH at the center of our galaxy as a lens. Here we  have considered that nature of black hole space-time is given by solutions of second order generalized Proca theories presented in Eq. (\ref{101}) and numerically estimated the values of different observables in Strong Field Limit for two separate lensing configuration namely a) when the source, lens and the observer is highly aligned with both the source and the observer is at infinity and b) taking the star S2 as a source.\\
We have first considered a lensing system where the source, lens and the observer are highly aligned and both the source and the observer are very far away from the lens. We numerically solved Eq. (\ref{123}) to get the radius of the photon sphere $r_{m}$ and angular radius of innermost image using the relation $\theta_{m}=u(r_{m})/r_{R}$. We have further assumed that the outer most image $\theta_{1}$ is resolved as a single image and all other images packed together at $\theta_{m}$. Then the observables- angular separation between the inner and outermost image $s$, the ratio of flux from the outermost image to those from all other image $\mathcal{R}$ and time delay between first and second relativistic image $\Delta T_{12}^{s}$ (formed on the same side of the lens) can be approximated by \cite{Ji:2013xua, Bozza:2002zj, Bozza:2003cp}
     \begin{equation}\label{151}
      \begin{aligned}
      s&=\theta_{1}-\theta_{m}\approx\theta_{m}\exp\left[\frac{\bar{b}_{rot}-2\pi}{\bar{a}_{rot}}\right]\\\mathcal{R}&=2.5\log_{10}\left(\dfrac{\mu_{1}}{\sum\limits_{n=2}^{\infty}\mu_{n}}\right)=\frac{5\pi}{\hat{a}_{rot}\ln 10}\\\Delta T^{s}_{12}&\approx2\pi \frac{\widetilde{a}_{rot}(a)}{\bar{a}_{rot}(a)}
      \end{aligned}
      \end{equation}
      So by measuring $\theta_{m}$, $s$ and $\mathcal{R}$ we can correctly predict strong lensing co-efficient $\bar{a}_{rot}$, $\bar{b}_{rot}$ and the minimum impact parameter $u_{m}$ and comparing them with the values predicted by a given theoretical model, we can identify the nature of the black hole. In Table-\ref{Tab3}, we have presented the numerical estimation of different observational parameters ($\theta_{m}$, $s$, $\mathcal{R}$, $\Delta T_{12}^{s}$). We also compare the results with those obtained from \KN (KN) black hole solution with charge $q$  whose line element can be expressed as \cite{1965JMP.....6..915N}.
	\begin{eqnarray}{\label{KN}}
	ds^{2}&=&-\left[1-\frac{2Mr}{\rho^{2}}+\frac{q^{2}}{\rho^{2}}\right]dt^{2}-\dfrac{4a \sin^{2}\vartheta}{\rho^{2}}\left[r-\frac{q^{2}}{2}\right]dt d\phi +\dfrac{\rho^{2}}{\Delta_{KN}}dr^{2}+\rho^{2}d\vartheta^{2}+ \nonumber\\
&&\left[r^{2}+a^{2}+\dfrac{2ra^{2}\sin^{2}\vartheta}{\rho^{2}}-\dfrac{ q^{2}a^{2}\sin^{2}\vartheta}{\rho^{2}}\right]\sin^{2}\vartheta d\phi^{2}
	 		\end{eqnarray} 
	 		where
	 		\begin{eqnarray}
& &{\Delta}_{KN}=r^{2}-2r+a^{2}+q^{2}\,,\qquad 
{\rho}^{2}=r^{2}+a^{2}\cos^{2}\vartheta\,. \nonumber 
\end{eqnarray}
     \begin{figure}
      	\centering
      	\begin{minipage}[b]{0.45\textwidth}
      		\includegraphics[width=\textwidth]{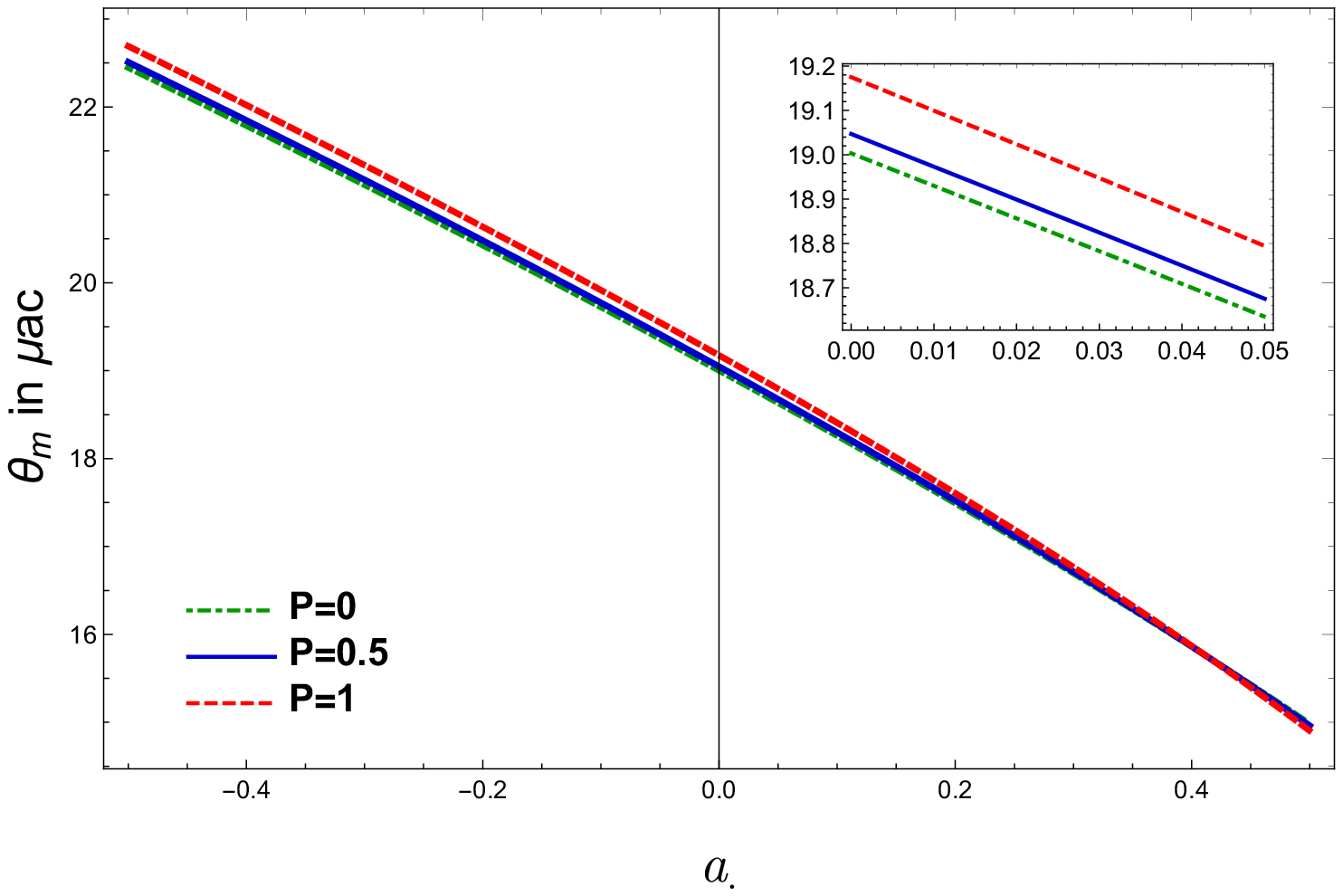}
      	\end{minipage}
      	\hfill
      	\begin{minipage}[b]{0.45\textwidth}
      		\includegraphics[width=\textwidth]{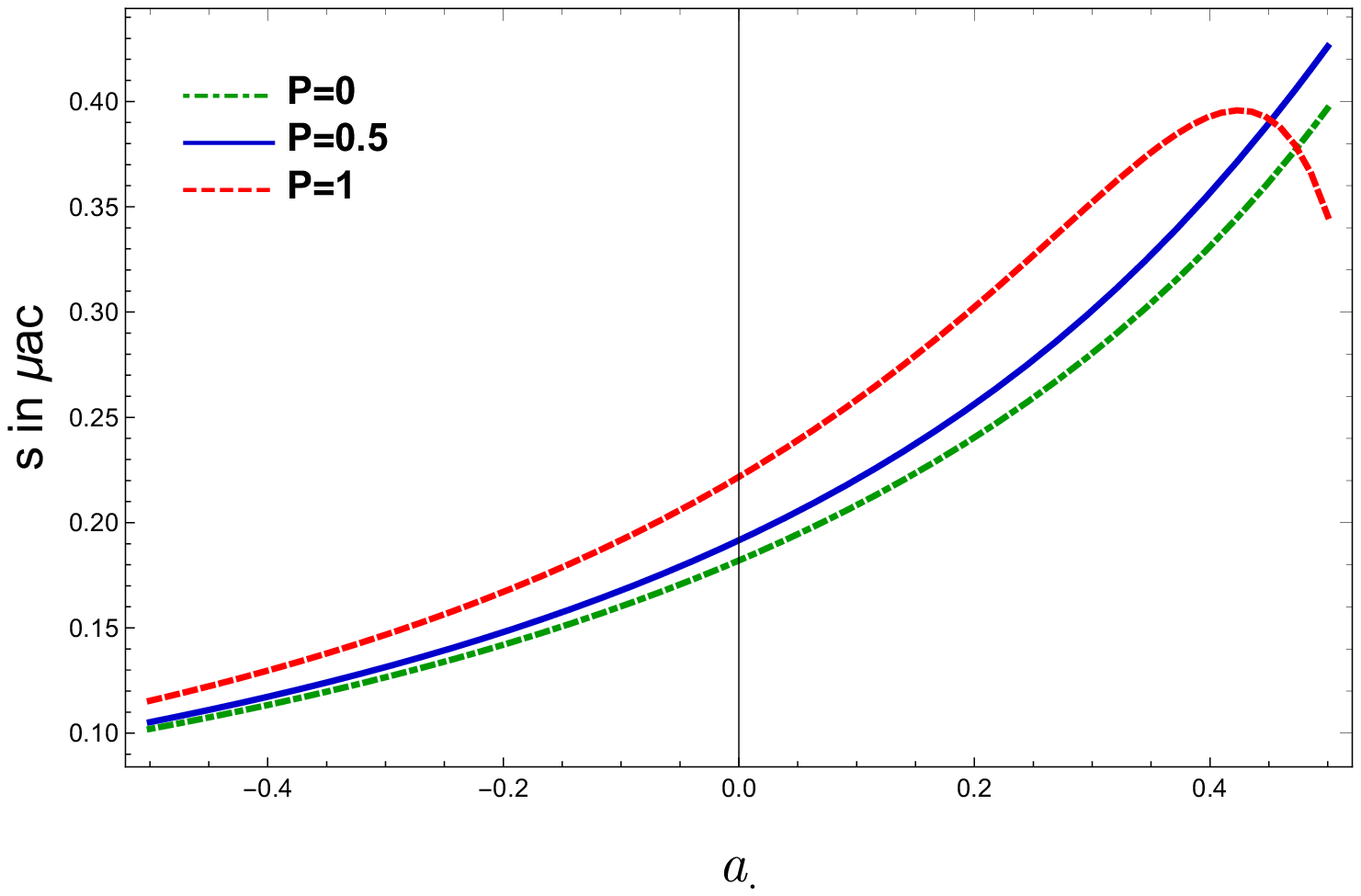}
      	\end{minipage}
      	\hfill
      	     	\begin{minipage}[b]{0.45\textwidth}
     		\includegraphics[width=\textwidth]{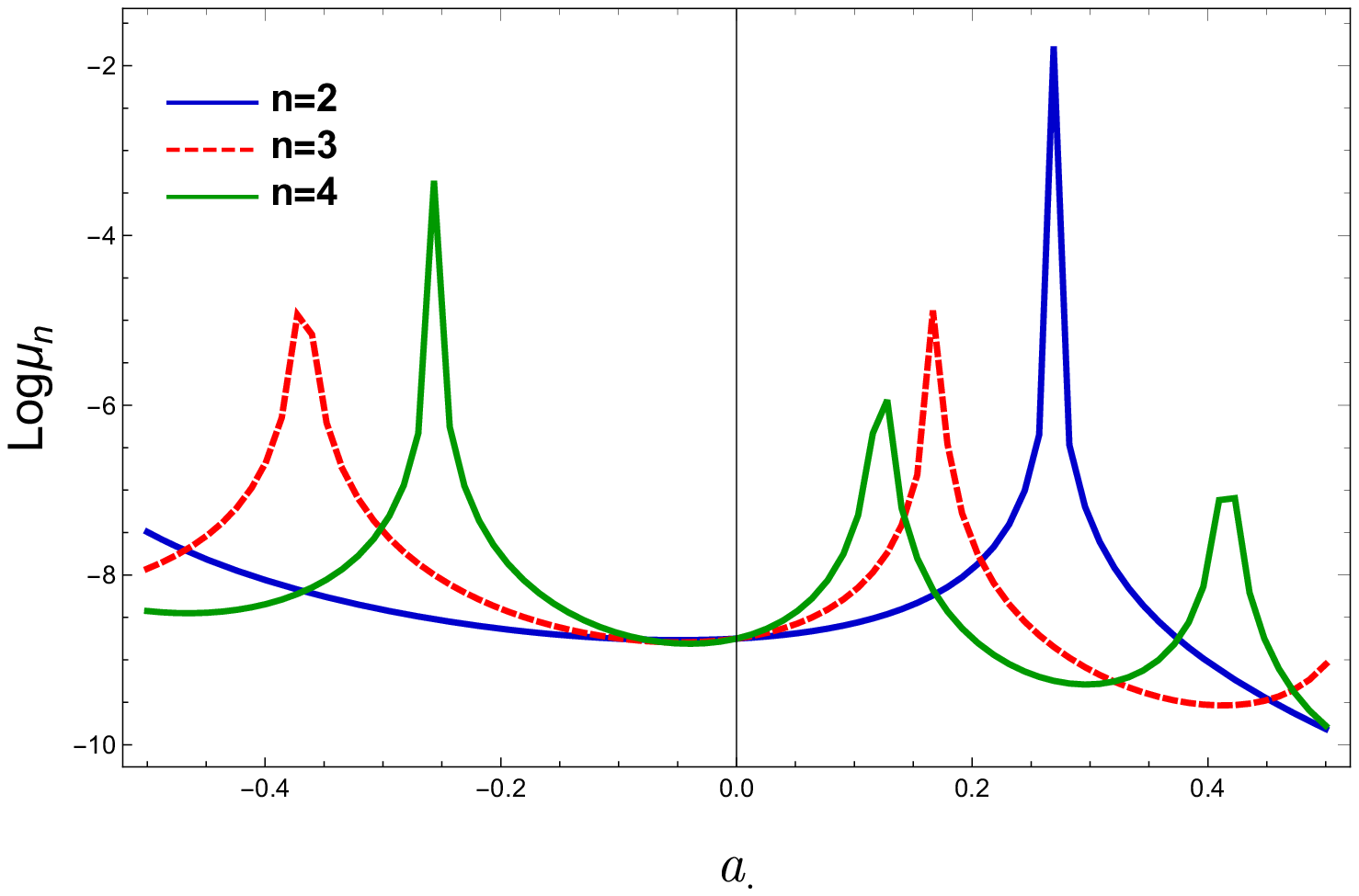}
     	\end{minipage}
      	\caption{Variation of different observables - (a) angular position of inner most image $\theta_{m}$, (top-left corner) and the (b) the angular difference between the outermost and innermost images s , (top-right corner)  for different values of Proca hair (c) Magnification of second, third and fourth relativistic images, $\log \mu_{n}$ (bottom) as a function of black hole spin $a$ have been presented.  Here we have taken S2 as source. Note that, the peaks in the magnification corresponds to caustic points. For drawing the caustics, we assume $P=0.5$; but the overall behavior remains the same for other values of $P$.}
          	\label{thetakvsa}
      \end{figure} 
In  Table-\ref{Tab3}, `$hair$' corresponds to Proca hair $P$ in the case of rotating black holes in generalized Proca theories ( Eq. (\ref{101})) and charged hair $q$ in the case of \KN (KN) black holes (Eq. (\ref{KN})). We also have plotted the observables as a function of hair parameter for these two black hole spacetime in Fig-\ref{observablesspin}. From Table-\ref{Tab3}, we can see that angular position of the inner most image $\theta_{m}$ decreases with the increase of black hole spin $a$. But it increases with the increase of the $hair$-parameter which is contrary to the KN case. Physically this implies that the size of the innermost Einstein ring is bigger for a slowly rotating black hole than those obtained from a rapidly rotating black hole. Moreover, that the size of the ring is bigger for the generalized Proca black holes space-time than those obtained from \KN space-time for the same value of the hair parameter. In Proca black hole spacetime, angular separation between the inner and outer most image increases with the increase of $a$. This angular separation increases monotonically with the increase of the $hair$-parameter similar to KN case. The relative flux $\mathcal{R}$ decreases with increase of both $a$ and $hair$-parameter similar to KN case.  Now the time delay between first and second relativistic image (formed on the same side of the lens) $\Delta T_{12}^{s}$ increases with the increase of $hair$-parameter which is in contrary to the KN case. Note that size of the inner-most Einstein ring  $\theta_{m}$ (or, the time delay between first and second relativistic image $\Delta T_{12}^{s}$) is maximum for the case $a,q=0$ (Schwarzschild black hole) for the black holes predicted by General Relativity. So any value of  $\theta_{m}$ ( or $\Delta T_{12}^{s}$) greater than those predicted in Schwarzschild space-time implies the existence of Proca hair. Thus by measuring the size of the innermost Einstein ring (or the time delay delay between first and second relativistic image), one can observationally verify the ``no-hair theorem"\cite{nohair}.\\
 Note that in Table-\ref{Tab3}, we have considered a highly aligned lensing system where both the source and observer are situated at infinity whereas in Tab-\ref{tab:tab2} we have taken S2 as a source. Now compare $a=0$ case (corresponds to non-rotating black hole with source at infinity) for different values of the $hair$-parameter presented in Table-\ref{Tab3} with the numerical values presented in Table-\ref{tab:tab2}. Here one can see the value of the angular separation between the innermost and outermost image $s$ is $\sim  \mathcal{O}(10)$ times higher for later case than those presented in Table-\ref{Tab3}. Thus the observable is in more detectable range when one considers S2 as a source. However, still, the small value of angular separation ($\sim  \mathcal{O}(10^{-1})$ $\mu$arc-sec) makes the detection of angular separation very difficult with present technologies. \\
     
      Now we will turn our attention to the gravitational lensing  by a rotating black hole in generalized Proca theories with S2 as a source. In this case a high alignment of the source with the optic axes does not happen due to inclination of the source orbit. So the simplified formula presented in Eq. (\ref{151}) does not work in this case. Rather we have to relay on  the more general formula of lensing given by Eq. (\ref{129}) and Eq. (\ref{147}) to have correct estimation of angular position and the magnification of the images. Using those expression, we have plotted the observables $\theta_{m}$ (top-left corner), $s$ (top-right corner) for different values of Proca hair $P$ and the magnification of the images corresponding to different winding number $n$ (bottom) as a function of $a$ in Fig--\ref{thetakvsa}. The peaks in the magnification corresponds to caustic points of the given lensing configuration where $\mathcal{K}(\gamma)$ vanishes (see Eq. (\ref{148})).  One can see that images corresponding to low winding number has fewer caustic points in the allowed range of $a$. As a result dimmer images meet caustic more often in the allowed range of $a$. For drawing the caustics, we assume $P=0.5$; but the overall behavior remains the same for other values of $P$.	      
 \section{Discussion and Conclusion} 
 In this paper, we discuss about different astrophysical aspects for a black hole in second order generalized Proca theories with derivative vector field interactions coupled to gravity. These black hole solutions are hairy and hence give us a perfect opportunity to observationally verify the "No-Hair theorem". We considered the super massive black hole in the center of galaxy is given by these generalized Proca theories and numerically estimated the values of different observables in strong field limit for two separate lensing configuration namely a) when the source, lens and the observer is highly aligned with both the source and the observer is at infinity and b) taking the star S2 as a source . For the latter case, we have shown that although the lensing system is not perfectly aligned, it gives observables in more detectable range . In early 2018, S2 was at its periapse position when one get maximum magnification \cite{Bozza:2004kq} and thus gave us a perfect opportunity to measure different lensing parameters in this time. \\
We also compare our results with those obtained from Reissner-Nordstr\"{o}m (Kerr-Newmann for rotating case)  black hole to see how the generalized Proca theory modifies the observables taking the stationary black holes predicted by GR as a reference.  Our study shows that the size innermost Einstein ring increases with the increase Proca hair $P$ for Proca black holes whereas the Einstein ring will shrink with the increase of charge $q$ for Reissner-Nordstr\"{o}m black hole (\KN for rotating case). The contribution of black hole spin can be well understood in the analysis of black hole shadow. Adding angular momentum to a black hole will cause a slight distortion in shape of the black hole shadow. Following Ref. \cite{Hioki:2009na}, we have shown that by measuring this distortion with respect to a reference circle, one can accurately measure the black hole parameters ($a, P$) . Thus by analyzing black hole shadow, we can directly probe black hole space time and hence the underlying gravity theory.  With the prospects of Even Horizon telescope as well as telescopes like SKA, one can resolve the black hole shadow with great accuracy and hence probing modified gravity through such observations is possible in near future. The other two observable angular separation between inner and outermost images and relative flux in between them exhibit same behavior as in the case of Reissner-Nordstr\"{o}m black hole (Kerr-Newmann for rotating case) with the change of hair parameter (which is Proca hair $P$ for Proca black holes and charged hair $q$ for \RN black hole). Angular separation between the images increases while the relative flux in between them decreases with the increase of the hair parameter. The time delay between first and second relativistic images, when both of them are formed on the same side of the lens, increases with the increase of $hair$-parameter which is in contrary to RN black hole (Kerr-Newmann for rotating case). The size of the inner-most Einstein ring  $\theta_{m}$ (or, the time delay between first and second relativistic image $\Delta T_{12}^{s}$) is maximum for $a,q=0$ case (Schwarzschild black hole) for the black holes predicted by General Relativity. So any value of  $\theta_{m}$ ( or $\Delta T_{12}^{s}$) greater than those predicted in Schwarzschild space-time, implies the existence of Proca hair. Thus by measuring the size of the innermost Einstein ring (or the time delay delay between first and second relativistic image), one can observationally verify the ``no-hair theorem".\\
Unfortunately, the angular separation between the inner and outermost relativistic image is extremely small ($\sim \mathcal{O}(10^{-1})$ $\mu$as while taking S2 as a source) which puts a great challenge for present technologies. However modern near-infrared (NIR) instruments like PRIMA \cite{PRIMA}, GRAVITY \cite{GRAVITY1}, ASTRA \cite{ASTRA} hope to achieve an astrometric accuracy of $10-100$ $\mu$as in combination with milli-arcsec angular-resolution imaging. With the help of these techniques, one can probe the Proca hair in near future. 

\begin{acknowledgements}
M.R. thanks INSPIRE-DST, Government of India for a Junior Research Fellowship.
\end{acknowledgements}

	\end{document}